\documentclass[longauth]{aa} % for the long lists of affiliations 
\usepackage{graphicx}
\usepackage{amssymb}
\usepackage{natbib}
\usepackage{multirow}
\usepackage{txfonts}
\usepackage{aas_macros}  % To allow the reading of ADS journal references in the bibliography
\bibpunct{(}{)}{;}{a}{}{,} % to follow the A&A style%%%%%%%%%%%%%%%%%%%%%%%%%%%%%%%%%%%%%%%%

\begin{document}

\title{Investigating the young Solar System analog HD\,95086
\thanks{Based on observations collected at the European
    Southern Observatory, Chile (ESO SPHERE Guaranteed Time Observation
Program 095.C-0273, 095.C-0298,  096.C-0241, 097.C-0865, 198.C-0209) and 
ESO HARPS Open Time Observation Program 099.C-0205, 192.C-0224.}}
\subtitle{A combined HARPS and SPHERE exploration}

\titlerunning{HARPS and SPHERE combined view of the young planetary architecture HD\,95086}
\authorrunning{Chauvin et al.}

\author{G. Chauvin \inst{1,2},
R. Gratton \inst{3},
M. Bonnefoy \inst{1},
A.-M. Lagrange \inst{1}, 
J. de Boer \inst{4},
A. Vigan \inst{5},   
H. Beust \inst{1},  
C. Lazzoni \inst{3}, 
A. Boccaletti \inst{6},
R. Galicher \inst{6},
S. Desidera \inst{3},
P. Delorme \inst{1},
M. Keppler \inst{7},
J. Lannier \inst{1},
A.-L. Maire \inst{7},
D. Mesa \inst{3},
N. Meunier \inst{1}, 
Q. Kral \inst{6},
T. Henning \inst{7}, 
F. Menard \inst{1},
A. Moor \inst{8}, 
H. Avenhaus \inst{9},
A. Bazzon\inst{9}, 
M. Janson \inst{7,10},
J.-L. Beuzit\inst{1}, 
T. Bhowmik\inst{6}, 
M. Bonavita\inst{11}, 
S. Borgniet\inst{6},
W. Brandner\inst{7}, 
A. Cheetham\inst{12}, 
M. Cudel\inst{1}, 
M. Feldt\inst{7},
C. Fontanive\inst{11}, 
C. Ginski\inst{13},
J. Hagelberg\inst{1}, 
P. Janin-Potiron\inst{14},
E. Lagadec\inst{14},
M. Langlois\inst{5,15},  
H. Le Coroller\inst{5},
S. Messina\inst{16}, 
M. Meyer\inst{8},  
D. Mouillet\inst{1}, 
S. Peretti\inst{12}, 
C. Perrot\inst{6}, 
L. Rodet\inst{1}, 
M. Samland\inst{7}, 
E. Sissa\inst{3},
J. Olofsson\inst{7,17},
G. Salter\inst{5},
T. Schmidt\inst{5},
A. Zurlo\inst{5,18},
J. Milli\inst{19},
R. van Boekel\inst{7},
S. Quanz\inst{8},
%P. A. Wilson\inst{20}, 
P. Feautrier\inst{1},
D. Le Mignant\inst{5},
D. Perret\inst{6},
J. Ramos\inst{7},
S. Rochat\inst{1}
}

\institute{
$^{1}$ Univ. Grenoble Alpes, CNRS, IPAG, F-38000 Grenoble, France. \\
$^{2}$ Unidad Mixta Internacional Franco-Chilena de Astronom\'{i}a, CNRS/INSU UMI 3386 and Departamento de Astronom\'{i}a, Universidad de Chile, Casilla 36-D, Santiago, Chile\\
$^{3}$ INAF - Osservatorio Astronomico di Padova, Vicolo dell’ Osservatorio 5, 35122, Padova, Italy\\
$^{4}$Leiden Observatory, Universiteit Leiden, PO Box 9513, NL-2300 RA Leiden, the Netherlands\\
$^{5}$ Aix Marseille Universit\'e, CNRS, LAM (Laboratoire d'Astrophysique de Marseille) UMR 7326, 13388 Marseille, France\\
$^{6}$ LESIA, Observatoire de Paris, PSL Research University, CNRS, Sorbonne Universités, UPMC Univ. Paris 06, Univ. Paris Diderot, Sorbonne Paris Cité, 5 place Jules Janssen, 92195 Meudon, France\\
$^{7}$ Max Planck Institute for Astronomy, K\"onigstuhl 17, D-69117 Heidelberg, Germany\\
$^{8}$ Konkoly Observatory, Research Centre for Astronomy and Earth Sciences, Hungarian Academy of Sciences, P.O. Box 67, H-1525 Budapest, Hungary\\
$^{9}$ Institute for Astronomy, ETH Zurich, Wolfgang-Pauli-Strasse 27, 8093 Zurich, Switzerland\\
$^{10}$  Department of Astronomy, Stockholm University, AlbaNova University Center, 106 91 Stockholm, Sweden\\
$^{11}$ SUPA, Institute for Astronomy, The University of Edinburgh, Royal Observatory, Blackford Hill, Edinburgh, EH9 3HJ, UK\\
$^{12}$  Geneva Observatory, University of Geneva, Chemin des Mailettes 51, 1290 Versoix, Switzerland\\
$^{13}$ Anton Pannekoek Institute for Astronomy, Science Park 904, NL-1098 XH Amsterdam, The Netherlands\\
$^{14}$  Universite Cote d’Azur, OCA, CNRS, Lagrange, France\\
$^{15}$ CRAL, UMR 5574, CNRS, Universit de Lyon, Ecole Normale Suprieure de Lyon, 46 Alle d'Italie, F-69364 Lyon Cedex 07, France\\
$^{16}$ INAF-Catania Astrophysical Observatory, via S.Sofia, 78 I-95123 Catania, Italy\\
$^{17}$ N\'ucleo Milenio Formaci\'on Planetaria - NPF, Universidad de Valpara\'iso, Av. Gran Breta\~na 1111, Valpara\'iso, Chile 
%Instituto de F\'isica y Astronom\'ia, Facultad de Ciencias, Universidad de Valpara\'iso, Av. Gran Breta\~na 1111, Valpara\'iso, Chile\\
$^{18}$ Núcleo de Astronomía, Facultad de Ingeniería, Universidad Diego Portales, Av. Ejercito 441, Santiago, Chile\\
$^{19}$ European Southern Observatory (ESO), Alonso de Córdova 3107, Vitacura, Casilla 19001, Santiago, Chile\\
$^{20}$ CNRS, UMR 7095, Institut d'Astrophysique de Paris, 98bis boulevard Arago, 75014, Paris, France\\
             \email{gael.chauvin@univ-grenoble-alpes.fr}
%             \thanks{The university of heaven temporarihd95086-sphere-v2-FME.pdfly does not
 %                    accept e-mails}
}

%   \date{Received September 15, 1996; accepted March 16, 1997}
% \abstract{}{}{}{}{} 
% 5 {} token are mandatory
  \abstract
      {HD\,95086 (A8V, 17\,Myr) hosts a rare planetary system for
        which a multi-belt debris disk and a giant planet of
        4--5~M$_{\rm{Jup}}$ have been directly imaged.}
      %% {In September 2013 was confirmed the discovery of young giant
      %%   planet orbiting the early-type star HD\,95086, member the
      %%   Lower Centaurus Crux association. Located at about 620~mas
      %%   (56~au) from the star, the planet luminosity is compatible
      %%   with a young 4--5~M$_{\rm{Jup}}$ mass planet revolving in a
      %%   two-belts debris disk architecture consisting of a warm inner
      %%   belt ($\le5~$au) and a cold outer disk ($60-100~$au)
      %%   previously revealed by \textit{Herschel} observations.}4-5~M$_{\rm{Jup}}$ at 56~au, A8V star
  % {} aims 
     {Our study aims to characterize the global architecture of this
       young system using the combination of radial
       velocity and direct imaging observations. We want to
       characterize the physical and orbital properties of
       HD\,95086\,b, search for additional planets at short and wide
       orbits and image the cold outer debris belt in
       scattered light.}
  % {} methods 
     {We used HARPS at the ESO 3.6\,m telescope to monitor the radial
     velocity of HD\,95086 over 2 years and investigate the existence of
     giant planets at less than 3~au orbital distance. With the IRDIS dual-band imager
     and the IFS integral field spectrograph of SPHERE at VLT, we
     imaged the faint circumstellar environment beyond $10$~au at six epochs between 2015 and
     2017.}
%% were to
%%        search for wide-orbit planets beyond $10~au$ and to follow-up
%%        HD\,95086\,b between 2015 and 2016. Together with past NaCo
%%        measurements properly re-calibrated, we have conducted an
%%        homogeneous analysis of NaCo and SPHERE data, that covers more
%%        than 4~yrs, to accurately monitor the planet position relative
%%        to the star. We used the combination of the HARPS and SPHERE
%%        detection limits to constrain the presence of additional
%%        giant planets in the system. Finally, we obtained differential polarimetric
%%        imaging in $J$-band with IRDIS to search for polarized light signal
%%        from the cold outer component of the HD\,95086 debris disk to
%%        explore the system architecture in terms of planet-disk
%%        connection.}
  % {} results
       {We do not detect additional giant planets around HD\,95086. We identified the nature
         (bound companion or background contaminant) of all point-like
         sources detected in the IRDIS field of
         view. None of them correspond to the ones recently discovered near the edge of the cold outer belt by ALMA. HD\,95086\,b is resolved for the first
         time in $J$-band with IFS. Its near-infrared spectral energy
         distribution is well fitted by a few dusty and/or young L7-L9
         dwarf spectral templates. The extremely red $1-4$\,$\mu$m spectral distribution
         is typical of
         low-gravity objects at the L/T spectral type transition. The planet's orbital motion is resolved
         between January 2015 and May 2017. Together with
         past NaCo measurements properly re-calibrated, our orbital
         fitting solutions favor a retrograde low to moderate-eccentricity orbit
         $e=0.2_{-0.2}^{+0.3}$, with a semi-major axis $\sim52$~au corresponding
         to orbital periods of $\sim288$\,yrs and an inclination that
         peaks at $i=141\,\degr$, which is compatible with a
         planet-disk coplanar configuration. Finally, we report the
         detection in polarimetric differential imaging of the cold outer
         debris belt between 100 and 300\,au, consistent in radial extent
         with recent ALMA 1.3\,mm resolved observations.}
%% close-in giant planets at less  than was
 %%           discovered with HARPS.  We derive
 %%           the most probable orbital solutions that best fit our
 %%           measurements using a Markov-Chain Monte Carlo Orbital
 %%           fitting approach. The solutions favor a retrograde
 %%           low-eccentricity orbit $e\la0.5$, with a semi-major axis
 %%           $\sim55$~au corresponding to orbital periods of
 %%           $\sim320$\,yrs and an inclination that peaks at
 %%           $i=120\pm20\,\degr$, which is compatible with a planet-disk
 %%           coplanar configuration. We do not detect any a
 {} 
 %    {}
   \maketitle
%
%------------------------------------------------------------------------
%------------------------------------------------------------------------

%%%%%%%%%%%%%%%%%%%%%%%%%%%%%%%%%%%%%%%%%%%%%%%%%%%%%%%%%%%%%%%%%%%%%%
\section{Introduction}

Dusty debris disks around pre- and main-sequence stars are signposts
for the existence of planetesimals and exoplanets
\citep{chen2012,marshall2014}. Numerous T Tauri and Herbig stars
indicate that the characteristic timescale for the dispersal of their primordial dusty, gaseous disks is a few million years
\citep{haisch2001,haisch2005,fedele2010}. Giant planet formation is expected to
play a key role in the evolution from protoplanetary to debris disks. This is indirectly confirmed by
submillimeter and near-infrared images of cool dusty debris
disks around main sequence stars \citep{schneider2014}. The presence of dust and the discovered disk
structures (ring, gap, warp and other asymmetries) could be indirect
indicators of the presence of giant planets. In that
context, it is striking to note that the majority of giant planets
recently discovered in imaging have been found around young, dusty,
early-type stars. This includes the breakthrough discoveries of HR\,8799 bcde (5-10~M$_{\rm{Jup}}$ at 10-64~au,
A8V star; \citealt{marois2008,marois2010}), $\beta$\,Pictoris\,b
(8-13~M$_{\rm{Jup}}$ at 9~au, A5V star; \citealt{lagrange2010}),
HD\,95086\,b (4-5~M$_{\rm{Jup}}$ at 56~au, A8V star;
\citealt{rameau2013a}) and 51\,Eri\,b (2~M$_{\rm{Jup}}$
at 14~au, F0V star; \citealt{macintosh2015}) with the exception of HIP65426\,b 
(6-12~M$_{\rm{Jup}}$ at 92~au, A2V star; \citealt{chauvin2017}) with no clear sign of debris disk in the system yet.

\begin{table}[t]
\caption[]{Physical properties of HD\,95086 from \cite{moor2013}
  for $T_{\rm{eff}}$, $log(g)$, [Fe/H], $v.sin(i)$, 
\cite{pecaut2012} for age, \cite{gaia2016} for distance,
  \cite{siess2000} for predicted mass
  and luminosity, \cite{cutri2003} and \cite{hog2000} for $J, H, K$ and $V$-band magnitudes.}
\label{prop}
\centering                          % used for centering table
\begin{tabular}{lll}
\noalign{\smallskip}\hline\noalign{\smallskip}
Parameter & Unit & Value \\
\noalign{\smallskip}\hline\noalign{\smallskip}
%\noalign{\smallskip}Star & &  \\
\noalign{\smallskip}$T_{\rm{eff}}$ & [K] & $7750\pm250$ \\
\noalign{\smallskip}$log(g)$ & [dex] & $4.0\pm0.5$ \\
\noalign{\smallskip}[Fe/H] &[dex]  & $-0.25\pm0.5$ \\
\noalign{\smallskip}$v.sin(i)$ & [km.s$^{-1}$] & $20\pm10$ \\
\noalign{\smallskip}distance & [pc] & $83.8\pm2.9^a$ \\
\noalign{\smallskip}Age & [Myr] &  $17\pm2$\\
\noalign{\smallskip}$L_{\star}$ & [$L_{\sun}$] & $5.7\pm1.7$ \\
\noalign{\smallskip}$M_{\star}$ & [$M_{\sun}$] & $1.6\pm0.1$ \\
\noalign{\smallskip}$V$ & [mag] & $7.36\pm0.01$ \\
\noalign{\smallskip}$J$ & [mag] & $6.906\pm0.019$ \\
\noalign{\smallskip}$H$ & [mag] & $6.867\pm0.047$ \\
\noalign{\smallskip}$K$ & [mag] & $6.789\pm0.021$ \\
\noalign{\smallskip}\hline\noalign{\smallskip}
%\noalign{\smallskip}Disk & &  \\
%\noalign{\smallskip}$i$ & (deg) &  $25\pm5$\\
%\noalign{\smallskip}PA & (deg) & $115\pm10$ \\
%\noalign{\smallskip}$e$ &  & $\le0.3$ \\
\end{tabular}
\tablefoot{($^a$) The Hipparcos new reduction from \cite{vanleeuwen2007} 
gives a significantly different distance of $90.4\pm3.4$\,pc.}
\end{table}

HD\,95086 belongs to this group of young planetary systems for
which a multi-belt debris disk and a giant planet have been directly
imaged. The central star is an early-type A8V kinematic member of the
Lower Centaurus Crux (LCC) association \citep{madsen2002,
  rizzuto2012}, located at $83.8\pm2.9$~pc \citep{gaia2016} and with an age
estimate of approximately 17~Myr (see main stellar properties in
Table~1). This star is known to harbor a debris disk with a high
fractional luminosity, 1.5$\times10^{-3}$, derived from IR and sub-mm
observations \citep{nilsson2010,rizzuto2012,chen2012}. \cite{moor2013}
did not detect any CO emission using APEX and set an upper limit
for the total CO mass between $1.4\times10^{-4}$ and
$1.7\times10^{-4}$~$M_{\oplus}$, excluding the possibility of an
evolved gaseous primordial disk (in agreement with \citealt{kral2017} secondary gas model predicting a CO mass content of $10^{-5}$~$M_{\oplus}$). The disk was resolved with
\textit{Herschel} PACS observations. \cite{moor2013} identified
in the deconvolved images at 70 and 100\,$\mu$m, a large disk structure
with a characteristic size fitted by an elliptical gaussian of
$540\times490$~au ($6.0\times5.4\,''$) with an inclination of
$25^o$ at a position angle of $105^o$. The modeling of the
spectral energy distribution (SED) combining \textit{2MASS},
\textit{IRAS}, \textit{WISE}, \textit{Spitzer}, \textit{Herschel}
photometry led them to favor a two-belt architecture with a
\textit{warm} inner belt ($187\pm26$\,K at $5.9\pm1.6~$au) and a colder outer
belt ($57\pm1.5$\,K at $63.7\pm4.4~$au) that would extend up to 270~au
to match the resolved PACS images. The revised values considering the new \textit{GAIA} DR1 distance 
of HD\,95086 remain within the previous uncertainty. Subsequent re-analysis by \cite{su2015} confirmed the presence of \textit{warm} and
\textit{cold} belts at the location derived by \cite{moor2013} using either
flat or gaussian disk models in their SED analysis. It also led them
to favor a three-component model with the presence of an additional
broad low-eccentricity ($e<0.3$) disk halo component extending up to
800~au to reconcile SED fitting and their new analysis of the PACS
far-infrared resolved images. They also suggest the presence of a closer
hotter belt at $\sim300$\,K in the terrestrial planet zone (that would
be located at $\sim2$~au) or the presence of a weak silicate feature
emitted by $\mu$m-size grains located in the \textit{warm} belt to explain the
excess of flux in the SED shortward of $10\,\mu$m. They also report the
detection of 69$\,\mu$m crystalline olivine feature from the disk not
spectrally resolved and contributing to $\sim5\,$\% of the total dust
mass. 

Very recently, \cite{su2017} resolved for the first time the \textit{cold} belt 
with ALMA 1.3\,mm observations obtained in January and April/May 2015. The disk 
emission is consistent with a broad ($\Delta R/R\sim0.84$), inclined ($30^o\pm3^o$) ring 
peaked at $200\pm6$\,au from the star. For a two-boundary model, the \textit{cold} belt is well described with sharp boundaries from $106\pm6$\,au 
to $320\pm20$\,au and a surface density distribution described by a power law with an 
index of $-0.5\pm0.3$. The deep ALMA map also reveals the presence of two sources near the edge of the \textit{cold} belt. The brightest one has however a constructed SED consistent with the one expected from a $z=2$ dusty galaxy. Finally, the disk multi-belt architecture led \cite{su2015,su2017} to discuss the existence of additional planets around HD\,95086 (single to multiple system) considering
the properties of HD\,95086\,b and dynamical constraints from the debris
disk distribution and the stability of multiplanetary system
configurations.

%The presence of dust and the disk structure
%(ring, gap, warp and other asymmetries) are indirect indicators of the
%presence of giant planets \citep{mouillet1997}. 

In 2012, in the course of a thermal angular differential imaging
survey of young, nearby intermediate-mass stars \citep{rameau2013b},
\cite{rameau2013a} observed the system with NaCo at the Very Large Telescope (VLT). They
discovered a very red candidate located at about 620~mas (51.9~au at 83.8\,pc) from
the star and with a luminosity compatible with a young
4--5~M$_{\rm{Jup}}$ planet. Subsequent observations in 2013 in
$K$-band and $L\,\!'$-band enabled to confirm the common proper
motion of the planetary candidate with HD\,95086 \citep{rameau2013b}. Further
characterization obtained with NaCo at VLT, NICI and during the early
science phase of GPI at Gemini-South confirmed the late-L type and
planetary nature of HD\,95086\,b
\citep{rameau2013c,meshkat2013,galicher2014}. More recently,
\cite{derosa2016} reported photometric $H$ ($1.5-1.8\mu$m) and
$K1$ ($1.9-2.2\mu$m) spectroscopic observations of HD\,95086\,b with
GPI. They confirm the L-type dusty atmosphere of this giant
planet evidenced by a featureless low-resolution spectrum and a
monotonically increasing pseudo-continuum in $K1$ consistent with a
cloudy atmosphere. In a complementary study, \cite{rameau2016} focused
their analysis on the system archictecture combining NaCo and GPI
observations between 2012 and 2016.  The orbital motion of the planet
was resolved. They report orbital solutions that favor, with 68\%
confidence, a semimajor axis of $61.7$~au (57.2 at 83.8\,pc) and an inclination of
$153.0\,\degr$ for eccentricities smaller than 0.21. They further
constrain the presence of inner planets in the system considering
the new GPI detection limit performances together with previous NaCo
$L\,\!'$-band observations.

%% Direct imaging is here an unique and viable technique to complete our
%% view of planetary system characteristics at wide orbits
%% ($\ge5$~au). This technique enables to directly study the planet-disk
%% connection to constrain the planet and disk physical properties,
%% evolution and formation. In the case of $\beta$\,Pictoris,
%% \citep{lagrange2012,chauvin2012} confirmed that $\beta$\,Pic\,b was
%% actually responsible for the disk innner warp geometry, pertubing the
%% planetesimals field and shaping the warp up to 40-60~au. HD\,95086 and
%% HR\,8799 share a common two-components architecture consisting of a
%% warm inner belt (175\,K at $8\pm2~$au), a cold outer belt (55\,K at
%% $60-190~$au) and a possible extended halo (up to 800\,au)
%% \citep{su2015}. \cite{kennedy2014} actually showed that the spectral
%% energy distributions of both systems are consistent with
%% two-temperature components compatible with dust emission arising from
%% two distinct radial locations.  Such an architecture would be
%% analogous to the outer Solar system’s configuration of Asteroid and
%% Kuiper belts separated by giant planets. In that context, the global
%% characterization of the young, planetary system around HD\,95086
%% combining HARPS and SPHERE measurements is particularly interesting.

\begin{table*}[t]
\caption{Obs Log of VLT/SPHERE observations}             % title of Table
\label{tab:obslog}
\centering
\begin{tabular}{llllllllll}     % 7 columns
\hline\hline\noalign{\smallskip}
UT Date    &    Instrument  &  Mode     & Filter   &  NDIT$\times$DIT$^a$  & $\rm{N}_{\rm{exp}}^a$      &   $\Delta\pi^a$     &  $\omega^a$    &   Strehl          & Airmass  \\ 
           &                &           &          &  (s)              &                       & ($\degr$)             & ($"$)     &  @1.6$\mu$m    &        \\
\noalign{\smallskip}\hline\noalign{\smallskip} 
03-02-2015 &    IRDIS       & DBI       &K1K2      & 4$\times$16       & 26                    & \multirow{2}{*}{17.5}   & \multirow{2}{*}{0.58 }      & \multirow{2}{*}{0.85 }     & \multirow{2}{*}{1.39}   \\
03-02-2015 &    IFS         & $R_\lambda=30$& YH     & 1$\times$64       & 26                    &               &        &       &   \\
\noalign{\smallskip} 
02-05-2015 &    IRDIS       & DPI       & J         & 2$\times$32     &  $76^c$                     & -                & 0.72       &  0.73       &  1.40 \\
\noalign{\smallskip}
%04-05-2015 &    IRDIS       & DBI       & H2H3      & 4$\times$64      & 16                    & 22.5             & XX       & 0.63     & 1.42 \\
%04-05-2015 &    IFS         &           & YJ        & XX$\times$XX     &                       & 22.5             & XX       & 0.63     & 1.42 \\
04-05-2015 &    IRDIS       & DBI       & K1K2      & 4$\times$64      & 13                     & \multirow{2}{*}{18.0 }            & \multirow{2}{*}{0.73 }      & \multirow{2}{*}{0.68}     & \multirow{2}{*}{1.40 }\\
04-05-2015 &    IFS         & $R_\lambda=30$ & YH    & 4$\times$64      & 13                     &             &       &     &  \\
%\noalign{\smallskip}
%11-05-2015 &    IRDIS       & DBI       & H2H3      & 15$\times$8      & 76                      & XX               & XX       &           &   \\
%11-05-2015 &    IFS         &           & YJ        & 64$\times$64     &                        & XX               & XX       &           &   \\
%\noalign{\smallskip}
\noalign{\smallskip}
18-01-2016$^b$ &    IRDIS       & DBI       & K1K2       & 5$\times$64      & 19                    & \multirow{2}{*}{28.4 }            & \multirow{2}{*}{1.56  }      &    \multirow{2}{*}{0.61  }      & \multirow{2}{*}{1.39}   \\
18-01-2016$^b$ &    IFS         & $R_\lambda=30$ & YH     & 5$\times$64      & 19                    &                &        &           &  \\
\noalign{\smallskip}
31-05-2016 &    IRDIS       & DBI       & K1K2       & 10$\times$64     & 7                     & \multirow{2}{*}{25.3 }              & \multirow{2}{*}{0.64  }    &  \multirow{2}{*}{0.82  }       & \multirow{2}{*}{1.41 }  \\
31-05-2016 &    IFS         & $R_\lambda=30$ & YH     & 10$\times$64     & 7                    &                &      &          & \\
\noalign{\smallskip}
10-05-2017 &    IRDIS       & DBI       & K1K2         & 10$\times$64     & 12                     & \multirow{2}{*}{36.6}              & \multirow{2}{*}{0.89}    &  \multirow{2}{*}{0.85}       & \multirow{2}{*}{1.39}  \\
10-05-2017 &    IFS         & $R_\lambda=30$ & YH     & 10$\times$64     & 12                    &                &      &          & \\
\noalign{\smallskip}
\hline\noalign{\smallskip}
\end{tabular}
\tablefoot{($^a$): NDIT refers to the number of integration per datacube, DIT to the integration 
time, $\rm{N}_{\rm{exp}}$ to the number of datacubes, $\Delta\pi$ to the parallactic 
angle variation during the sequence and $\omega$ to the seeing conditions. ($^b$): January 18th, 2016 
observations suffered from low-wind effect as illustrated in the Zoom-in image of Fig.\,\ref{fig:image}. ($^c$): For DPI, note that one HWP or polarimetric cycles is composed of four data cubes. Twenty cycles were obtained for HD\,95086 leading to a total of 76 exposures as one poor-quality cycle was discarded.}
\end{table*}

Despite the increasing interest for this system in the past years,
several fundamental questions remain unanswered regarding the origin
and architecture of the debris disk (owing to the limited amount of
spatial information), its connection to the presence of HD\,95086\,b
and additional planets in the system or the formation and the physical
properties of HD\,95086\,b itself. To further characterize this young
planetary system, we present in this paper a combined exploration
using HARPS and SPHERE observations. We aim at exploring the global
environment of HD\,95086 including the giant planet HD\,95086\,b, the
presence of additional planets from a few stellar radii up to 800~au
and the debris disk architecture. We report in Sect.~2 the observing
set-up and strategies, as well as the data reduction with both
instruments. In Sect.~3, we report the study of all detected
point-sources with SPHERE in the close vicinity ($\sim10-800$\,au) of
HD\,95086. In Sect.~4, we present the combined astrometric results
obtained with NaCo and SPHERE to derive the best orbital solution for
the planet. In Sect.~5, we report the first detection of HD\,95086\,b
in $J$-band and analyze the spectral energy distribution of the planet
together with the published photometry and spectrum from the
literature. In Sect.~6, we combine HARPS and SPHERE detection limits
to constrain the presence of additional giant planets in that system
at short and long periods.  In Sect.~7, we report the low-S/N
detection of the cold outer disk in polarized intensity located at a
distance consistent with the debris disk architecture reported by the
analysis of the SED and \textit{Herschel} far-IR images, and the recent ALMA 1.3\,mm resolved observations. In Sect.~8, we finally discuss the
global system architecture in the view of our results to investigate
the origin of the inner belt and the presence of additional planets in
viable dynamical configurations.

%%%%%%%%%%%%%%%%%%%%%%%%%%%%%%%%%%%%%%%%%%%%%%%%%%%%%%%%%%%%%%%%%%%%%%
\section{Observations and data reduction}

%%%%%%%%%%%%%%
\subsection{VLT/SPHERE Guaranteed time observations}

The SPHERE planet-finder instrument installed at the VLT
\citep{beuzit2008} is a highly specialized instrument, dedicated to
high-contrast imaging and spectroscopy of young giant exoplanets. It
is based on the SAXO extreme adaptive optics system \citep{fusco2006,
  petit2014, sauvage2010}, which controls a deformable mirror with 
$41\times41$ actuators, and 4 control loops (fast visible tip-tilt,
high-orders, near-infrared differential tip-tilt and pupil
stabilization).  The common path optics employ several stress polished
toric mirrors \citep{hugot2012} to transport the beam to the
coronagraphs and scientific instruments. Several types of
coronagraphic devices for stellar diffraction suppression are
provided, including apodized pupil Lyot coronagraphs (Soummer 2005)
and achromatic four-quadrants phase masks \citep{boccaletti2008}. The
instrument has three science subsystems: the infrared dual-band imager
and spectrograph (IRDIS, \citealt{dohlen2008}), an integral field
spectrograph (IFS; \citealt{claudi2008}) and the Zimpol rapid-switching
imaging polarimeter (ZIMPOL; \citealt{thalmann2008}).

\begin{figure}[t]
\hspace{0.2cm}
\includegraphics[width=\columnwidth]{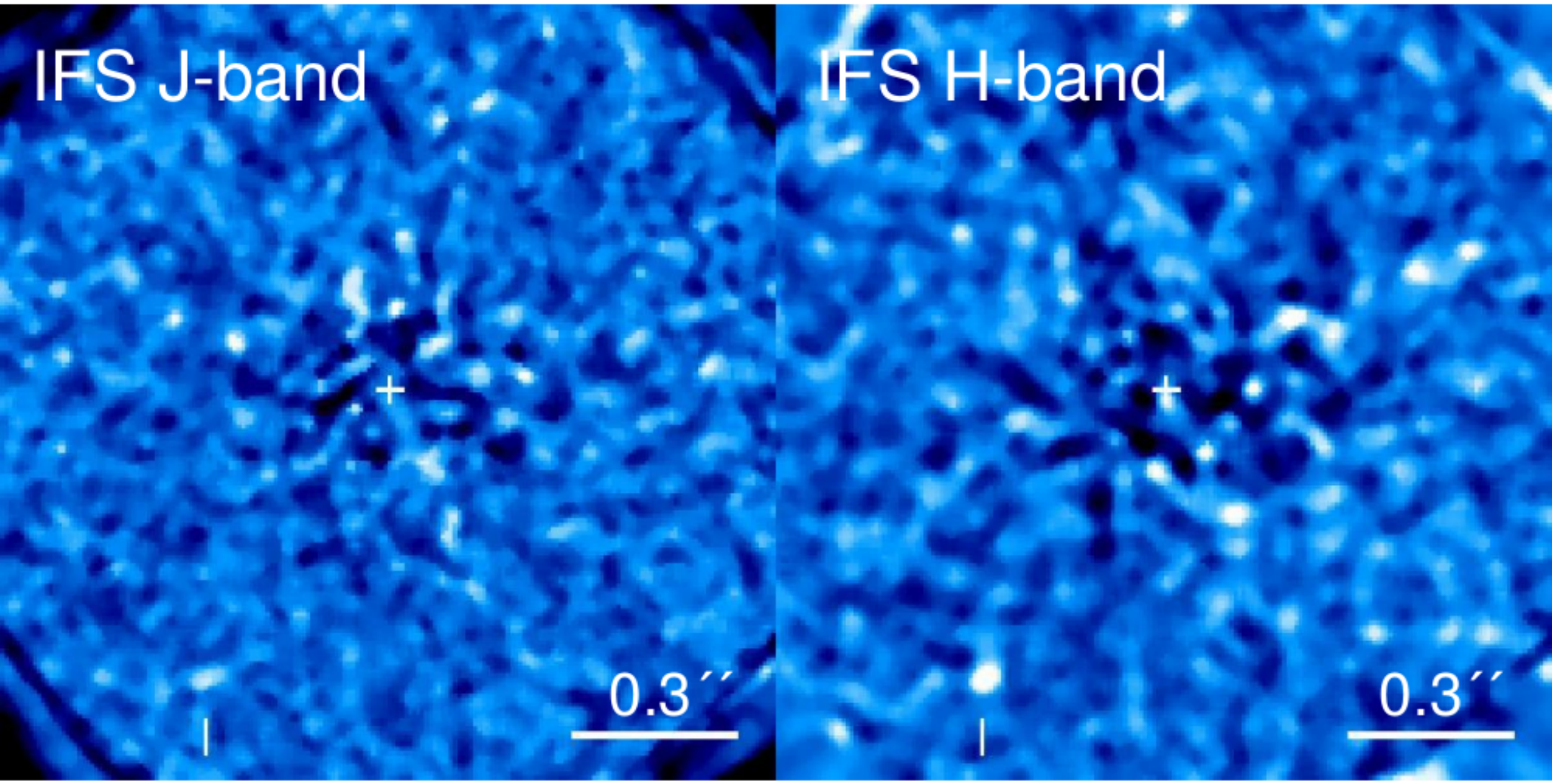}
\begin{centering}

\caption{\textit{Left:} IFS J-band image of HD\,95086 from a spectral
  PCA analysis of the combined IFS-YH datacubes obtained in February
  2015, May 2015, January 2016 and May 2016. The planet is not
  detected in the individual epochs and marginally with a S/N of
  $\sim3$ considering the combination of all epochs. \textit{Right:}
  Same for $H$-band. The planet is detected with a S/N of $\sim5$
  considering the combination of all epochs.}

\label{fig:ifs}
\end{centering}
\end{figure}

\begin{figure*}[t]
\hspace{0.2cm}
\includegraphics[width=18cm]{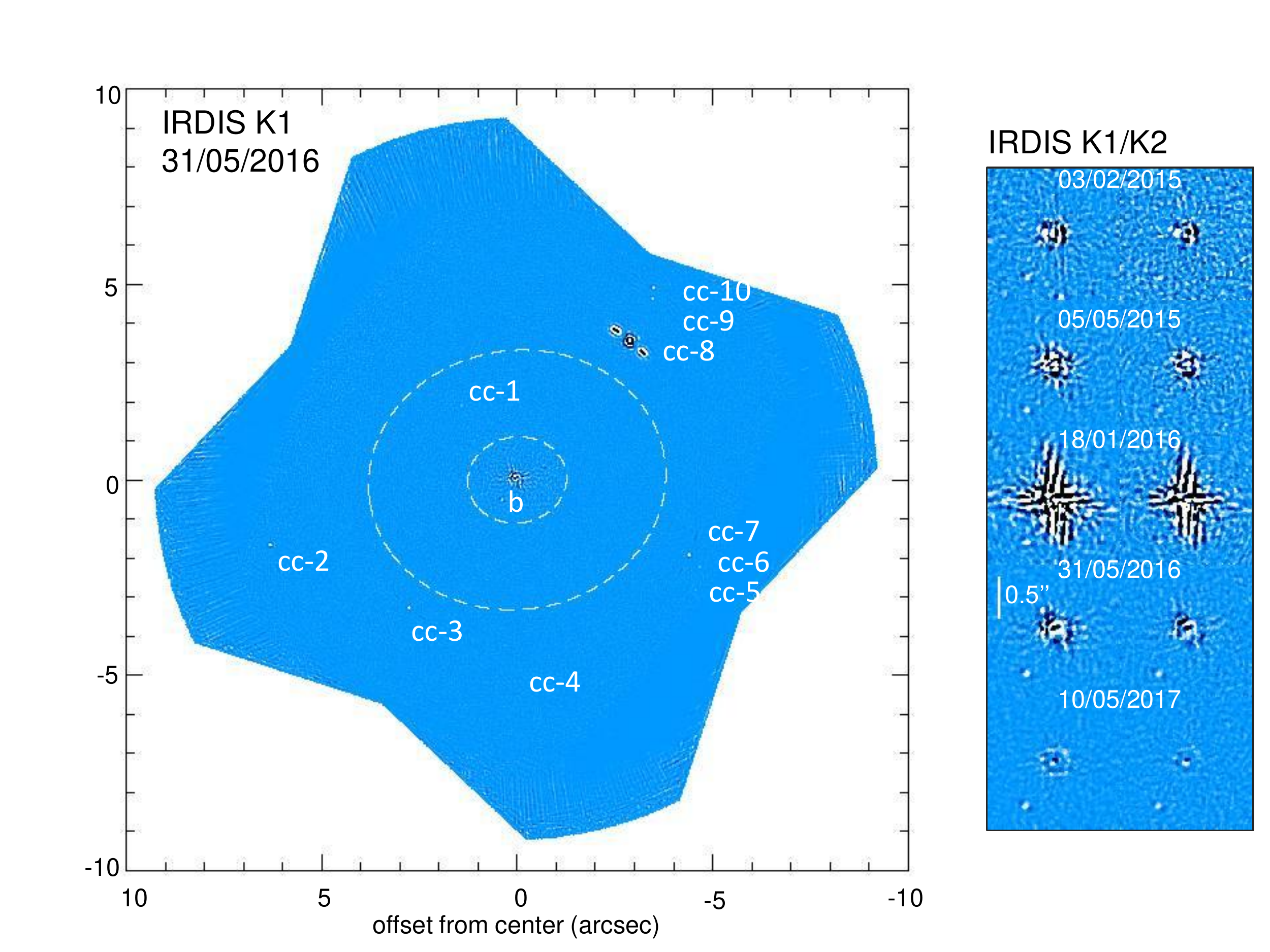}
\begin{centering}
\caption{\textit{Left:} IRDIS $K1$-band full-combined image of
  HD\,95086 from May 31st, 2016. All companion candidates have been
  marked. HD\,95086\,b is well detected at a separation of
  $622\pm3$~mas and position angle of $147.5\pm0.3$~deg from
  HD\,95086. The sharp boundaries at 106 and 320\,au of the \textit{cold} outer belt resolved by ALMA are reported in the IRDIFS FoV (dashed line).
\textit{Right:} Zoom-in of IRDIS $K1$ and $K2$-band
  images of HD\,95086 and HD\,95086\,b observed in February 3rd, 2015,
  May 5th, 2015, January 18th, 2016, May 31st, 2016, and May 10th, 2017. January 18th, 2016 observations 
suffered from low-wind effect as shown by the increased level of residuals.}
\label{fig:image}
\end{centering}
\end{figure*}

\subsubsection{Simultaneous integral field spectroscopy and dual-band imaging}

As part of the SHINE (SpHere INfrared survey for Exoplanets) GTO
campaign (095.C-0298, 096.C-0241, 097.C-0865, and 198.C-0209), aimed at the detection
and characterization of extrasolar planets in the near infrared, HD\,95086 was observed at
five epochs on February 3rd, 2015, May 5th, 2015, January 18th, 2016,
May 31st, 2016, and May 10th, 2017. The data were acquired in IRDIFS-EXT mode, using IRDIS
in dual-band imaging (DBI, \citealt{vigan2010}) mode with the $K_1K_2$ filters ($\lambda_{K_1} =
2.1025 \pm 0.1020,\mu$m - $\lambda_{K_2} = 2.2550 \pm 0.1090,\mu$m), and
IFS in the $Y-H$ ($0.97-1.66\,\mu$m) mode in pupil-tracking. This
combination enables the use of angular and/or
spectral differential imaging techniques to improve the contrast perfomances at the
sub-arcsecond level.

The standard SHINE observing sequence is composed of: one PSF
sub-sequence registered with a series of off-axis
unsaturated images obtained with an offset of $\sim0.4\,''$ relative
to the coronagraph center (produced by the Tip-Tilt mirror) and a neutral density (here ND1.0, which
reduces the flux by a factor
∼10 to avoid saturation)\footnote{https://www.eso.org/sci/facilities/paranal/instruments}. During this observation, the AO visible tip-tilt and high-order loops remain
closed to provide a diffraction-limited PSF. This sub-sequence is
followed by a “star center” coronagraphic observation
where four symmetric satellite spots are created by introducing a
periodic modulation on the deformable mirror. They are produced by the
SAXO high-order deformable mirror which creates a bi-dimensional wave
and are located at a separation of $14.2\times\lambda/D$ with adjustable
intensity and for two hardcoded configurations
(vertically/horizontally aligned as a "cross" or along the
"diagonals"; see \citealt{langlois2013}). The four satellite spots enable
an accurate determination of the star position behind the coronagraphic
mask for the following deep coronagraphic sequence. We used here the
smallest apodized Lyot coronagraph (ALC-YH-S) with a focal-plane mask
of 185~mas in diameter. The deep coronagraphic sub-sequence that follows typically lasts 1.5-2.0~hrs. The full sequence
is then concluded with a new “star center” sequence, a new PSF
registration, as well as a short sky observing sequence 
for fine correction of the hot pixel variation during the night. For February 3rd, 2015, May 31st, 2016, and May 10th, 2017 the deep
coronagraphic sequence were continuously obtained with the
four satellite spots to accurately control the stellar position. The observing settings
 and conditions of all epochs are
compiled in Table~\ref{tab:obslog}. January 18th, 2016 observations 
suffered from low-wind effect \citep{sauvage2016} as illustrated in the Zoom-in 
image of Fig.\,\ref{fig:image} degrading the XAO correction and SPHERE performances.

\begin{figure*}[t]
\hspace{0.2cm}
\includegraphics[width=\columnwidth]{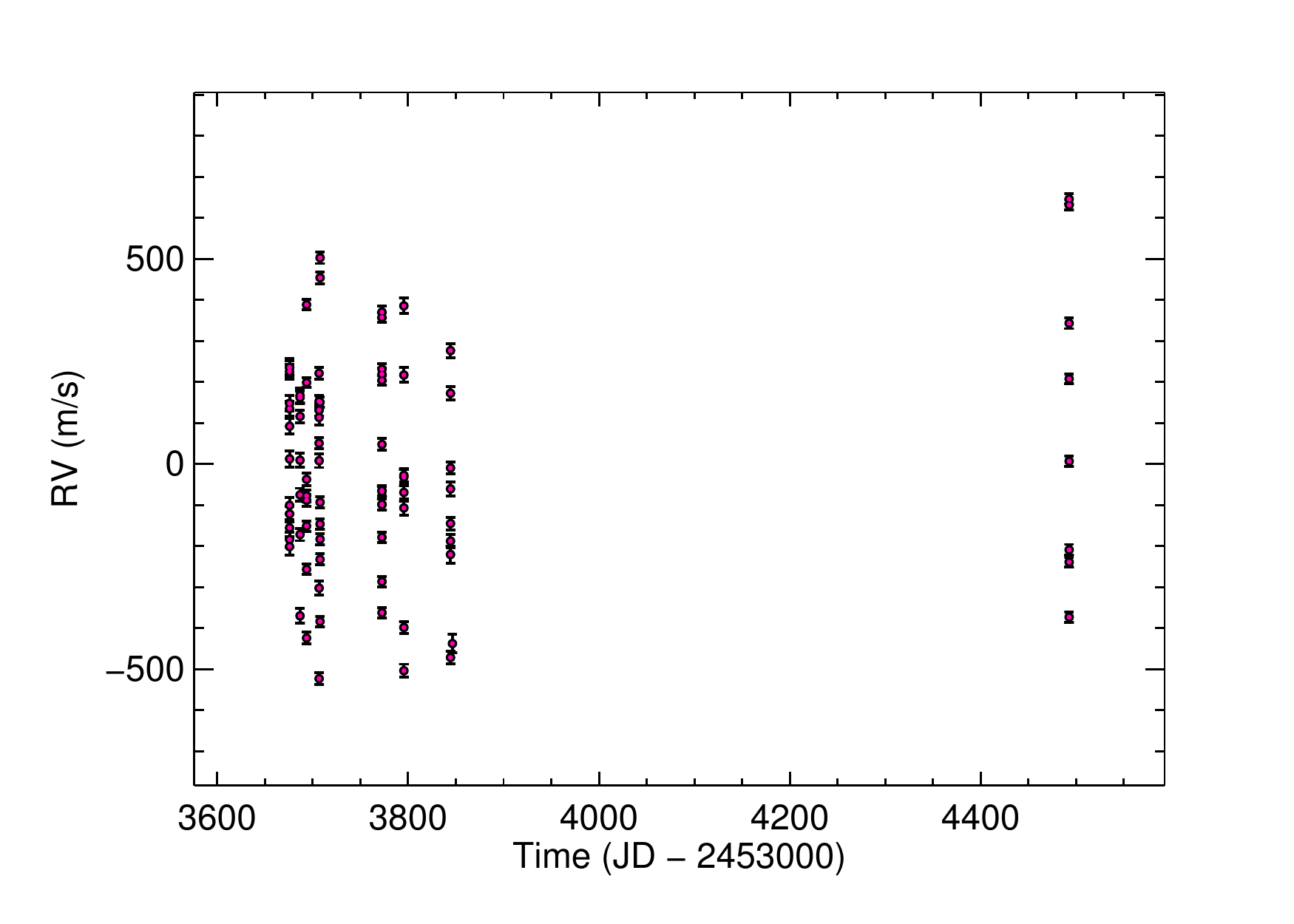}
\includegraphics[width=\columnwidth]{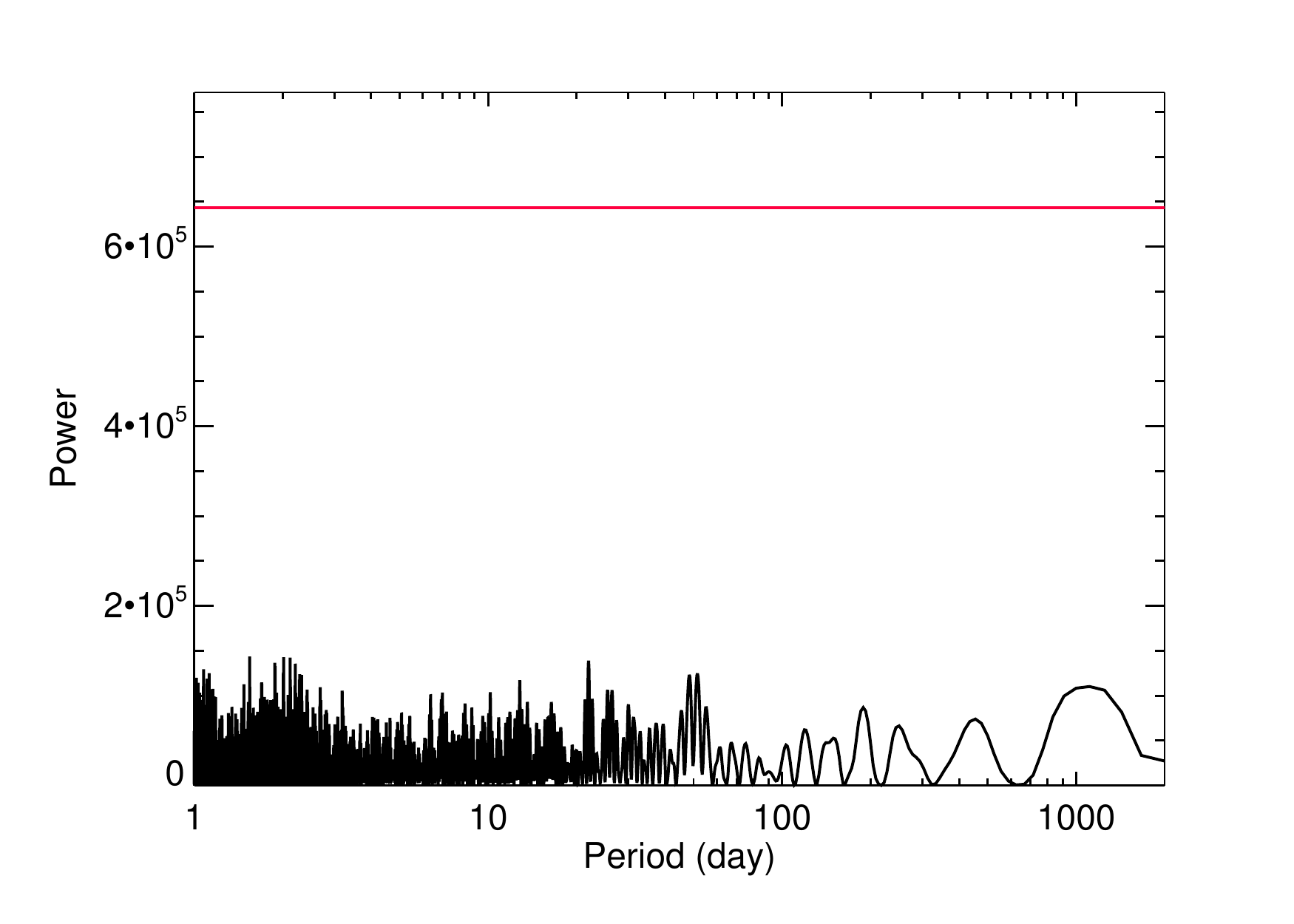}
\begin{centering}
\caption{\textit{Left:} HD\,95086 radial velocity variations observed with HARPS with error
  bars. \textit{Right:} Periodogram of HD\,95086 radial velocities. The false alarm
  probabilities at 10\% is drawn as a red line. There is no clear detection of significant periodic radial velocity variation in our data.}
\label{fig:rvharps}
\end{centering}
\end{figure*}

To calibrate the IRDIS and IFS dataset on sky, the astrometric field
47\,Tuc was observed. The platescale and True North solution at each
epoch is reported in Table~\ref{tab:results} based on the long-term
analysis of the GTO astrometric calibration described by
\cite{maire2016}. The rotation correction considered to align images
to the detector vertical in pupil-tracking observations is
$-135.99\pm0.11\degr$. Anamorphism correction is obtained by
stretching the image $Y$-direction with a factor of
$1.0060\pm0.0002$. All IRDIS and IFS datasets were reduced using the
SPHERE Data Reduction and Handling (DRH) automated pipeline
\citep{pavlov2008} at the SPHERE Data Center (SPHERE-DC) to correct for each
datacube for bad pixels, dark current, flat field and sky
background. After combining all datacubes with an adequate calculation of
the parallactic angle for each individual frame of the deep
coronagraphic sequence, all frames are shifted at the
position of the stellar centroid calculated from the initial star
center position. For May 5th, 2015, and January 18th, 2016,
the determination of the stellar position before and after the deep
coronagraphic was used to control the centering stability. A centroid
variation of $0.2$~pixels at both epochs was derived and quadratically
added for the final astrometric budget at each specific epoch. For the
February 3rd, 2015, May 31st, 2016, and May 10th, 2017 datasets, we took advantage of the
waffle-spot registration to apply a frame-to-frame recentering. The SPHERE-DC corrected products were used as input to the SHINE 
\textit{Specal} pipeline which applies flux normalization including the coronagraph transmission correction, followed
by different angular and spectral differential imaging
algorithms (Galicher et al. 2018, in prep). After a Fourier spatial filtering 
removing low-spatial frequencies, 
the TLOCI \citep{marois2014} and PCA \citep{soummer2012} algorithms
were specifically applied to both IRDIS and IFS data. To attenuate the signal, the Specal TLOCI implementation locally subtracts the stellar speckle pattern for each frame in annuli of $1.5\times\textit{FWHM}$ further divided in sectors. The subtraction is based on a linear combination of the best 20 ($N$ parameter) correlated reference images calculated in the optimization region and selected to minimize the self-subtraction at maximum $20\%$ ($\tau$ parameter). Please refer to \citep{galicher2011} and \citep{marois2014} for further description of the reference frame selection, and the subtraction and optimization regions. For IRDIS, the Specal PCA uses each frame subtracted from its average over the field of view to estimate the principal components. The first 5 components are considered for the final subtraction. For IFS, the spectral diversity is in addition exploited after proper rescaling and renormalization of the IFS datacubes as detailed by \cite{mesa2015}. The first 100 principal components are subtracted. Alternatively to Specal, the 
IPAG-ADI pipeline \citep{chauvin2012}, used to extract the early NaCo astrometric and photometric measurements of HD\,95086\,b \citep{rameau2013b,rameau2013a}, was run with the sADI and PCA algorithms to obtain a
consistency check. Astrometric, photometric and detection limit results were estimated using injected fake planets 
and planetary signature templates to take into account any biases related to the
data processing. The full FoV IRDIS image of May 31st, 2016 is shown in Fig.~\ref{fig:image} together with the IRDIS 
$K_1$ and $K_2$ sub-images of HD\,95086\,b obtained at each epoch.

\subsubsection{Polarimetric differential imaging}

On May 2nd, 2015, as part of the DISK GTO program (095.C-0273), a standard
differential polarimetric imaging (DPI, \citealt{langlois2014}) sequence of HD\,95086 was
obtained in $J$-band to possibly detect and study the spatial
distribution of small dust grains in the disk surface layers through
their scattered light \citep{kuhn2001}. The
debris disk around that source has never been resolved at visible or
near-infrared wavelengths. The small Apodized-Lyot coronagraph
(ALC-YJ-S) was used with a focal-plane mask of 145~mas in diameter. In
DPI mode, IRDIS provides two beams, in which wire-grid polarizers are
inserted, and lead to ordinary and extraordinary polarization
states. The half-wave plate (HWP) that controls the orientation of the
polarization is set to four positions shifted by $22.5\,\degr$ in order to
construct the classical set of linear Stokes vectors. A total of 76
datacubes of 2 frames were acquired for a total exposure time of
49~min on source in field stabilized mode. The detail of the observing setup, airmass,
parallactic angle variations are also reported in
Table~\ref{tab:obslog}.

Two pipelines developed for high-contrast differential polarimetric imaging were used to calculate the Stokes parameters Q and
U. The first one is based on the double-ratio method \citep{avenhaus2014}. A complete description is provided by 
\cite{benisty2015}. The second one is based on the double-difference approach and is described by \cite{deboer2016}. 
Both methods were independently applied 
to the HD\,95086's dataset. Since the scattered light from a circumstellar
disk is expected to be linearly polarized in the azimuthal direction
under the assumption of single scattering, it is beneficial to
describe the polarization vector field in polar rather than Cartesian
coordinates. Once data have been corrected from distortion, True North
and instrumental effects such as the angular misalignment of the HWP,
the Q$_\phi$, U$_\phi$ radial polarized Stokes parameters are
calculated following the principle of \cite{schmid2006}.  In this
coordinate system, the azimuthally polarized flux from a circumstellar
disk appears as a consistently positive signal in the Q$_\phi$ image,
whereas the U$_\phi$ image remains free of disk signal and provides a
convenient estimate of the residual noise in the Q$_\phi$ image
\citep{schmid2006}. This assumption is valid for low-inclination 
optically-thin debris disks \citep{canovas2015}. The outcome of the 
observations are discussed in
Sec.~\ref{sect:disk}.

%%%%%%%%%%%%%%
\subsection{ESO 3.6\,m telescope HARPS radial velocity}

Between 2014 and 2016, high-signal-to-noise spectra of the A8V early-type star HD\,95086 were obtained with
the HARPS spectrograph \citep{pepe2002} installed on the 3.6\,m ESO
telescope at La Silla Observatory (Chile) in the southern
hemisphere. These observations were part of two HARPS Open Time and Large Programs (099.C-0205 and 192.C-0224) focused on
the search for exoplanets around young, nearby stars. The radial
velocities were measured using the dedicated SAFIR code \citep{galland2005} based on the
Fourier interspectrum method developed by \citep{chelli2000}. A reference spectrum (average of all spectra)
is used instead of a classical binary mask for cross-correlation and
estimation of the radial velocity data. It offers the possibility to apply a differential approach based on 
star spectrum itself and to use the low-frequency structures and spectral discontinuities to estimate the best value of the stellar velocity in presence of stellar noise. The application of the Fourier approach to early-type stars that present a small number of stellar lines, usually broadened and blended by stellar rotation, has proven to be very succesful for the detection of planets \cite[e.g.][]{borgniet2017}. The results of the radial
velocity variation and periodogram are reported in
Fig.\,\ref{fig:rvharps} and cover a timeline of more than
1000~days. The typical uncertainty associated with HARPS data is of
15~m.s$^{-1}$ on average. The observed scatter in the radial velocity
variation is one order of magnitude larger than the individual radial
velocity uncertainty and is probably due to the presence of pulsations
as expected for young, early-type stars with similar spectral types than HD\,95086.

\begin{figure}[t]
\hspace{0.2cm}
\includegraphics[width=8.5cm]{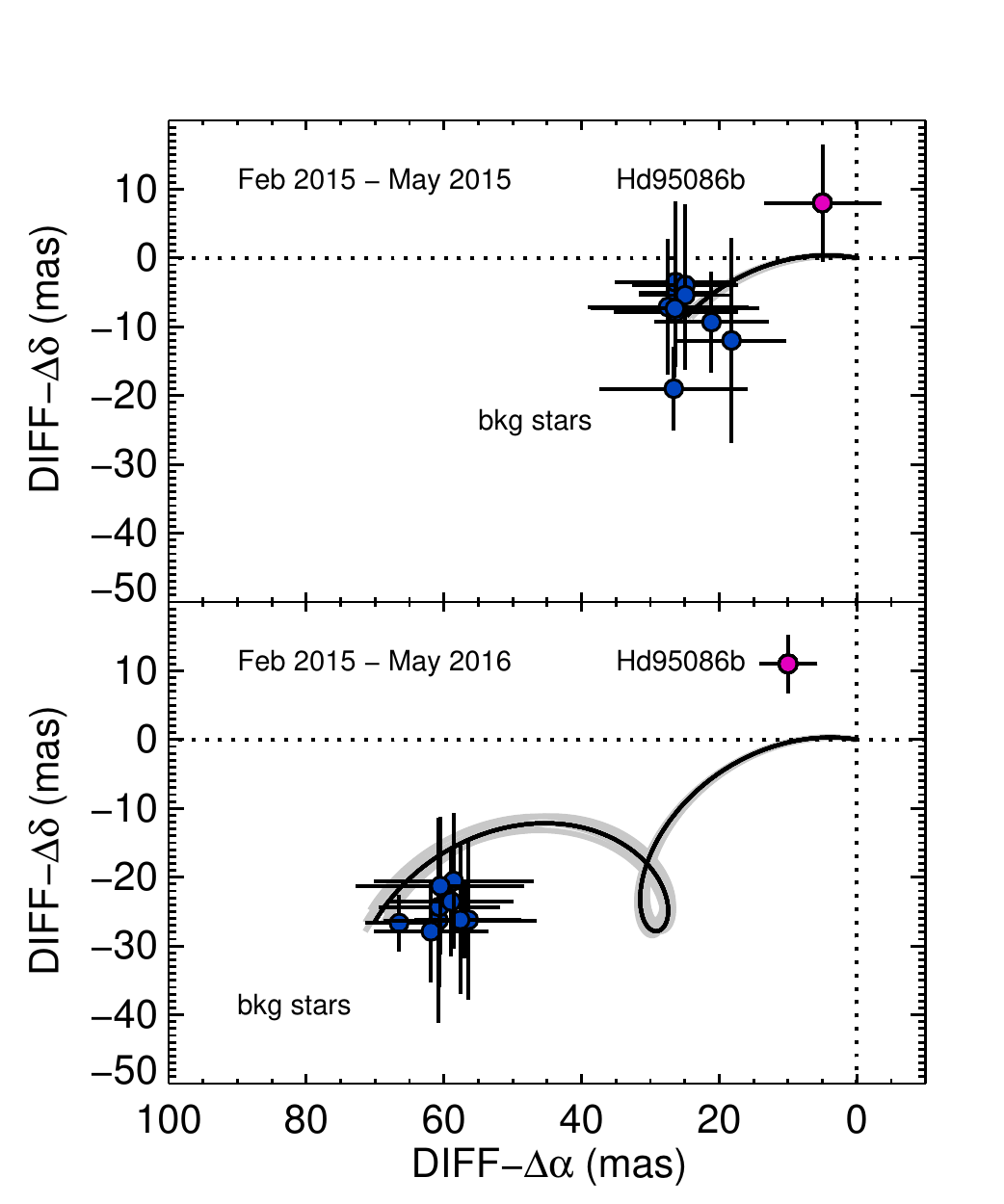}
\begin{centering}
\caption{\textit{Top}: SPHERE/IRDIS relative astrometry differences
  (\textit{dark blue filled circles} with uncertainties) of the offset
  positions of all point-like sources detected in the close vicinity
  of HD\,95086 (see Fig.~\ref{fig:image}) between May 5th, 2015, and
  Feb 3rd, 2015. The expected variation of offset positions, if the
  candidates are stationary background objects is shown (\textit{solid line} with
  uncertainties). The variation is estimated based on the parallactic
  and proper motions of the primary star, as well as the initial
  offset position of the companion candidates from HD\,95086. The planet position is indicated with a
\textit{pink} circle. Except
  HD\,95086\,b, all companion candidates are clearly identified as
  background contaminants. \textit{Bottom}: Same as top panel but between May 31st,
  2016, and Feb 3rd, 2015.}
\label{fig:ppmtest}
\end{centering}
\end{figure}

%%%%%%%%%%%%%%%%%%%%%%%%%%%%%%%%%%%%%%%%%%%%%%%%%%%%%%%%%%%%%%%%%%%%%%
\section{Point-source identification in the stellar vicinity}
\label{sect:point}

\begin{table*}[t]
\caption{IRDIS Relative astrometry and photometry of HD\,95086 A and b}             % title of Table
\label{tab:results}
\centering
\begin{tabular}{llllllllll}     % 7 columns
\hline\hline\noalign{\smallskip}
UT Date    &    Ins-Filter          &  $\Delta\alpha$     & $\Delta\delta$      & Sep.         &   PA                &  Contrast           &     True North             &    Platescale     & Ref. \\
           &                      &  (mas)              &   (mas)             &  (mas)             &   ($\degr$)             &  (mag)              &     (deg)          & (mas)             &  \\ 
\noalign{\smallskip}\hline\noalign{\smallskip} 
12-01-2012 &    NaCo-$L\!'$       &  $294\pm8$          &  $-550\pm8$         &  $624\pm8$         &  $151.9\pm0.8$      & $9.8 \pm0.4$         &   $-0.57\pm0.10$   & $27.11\pm0.06$   & 1              \\ 
14-03-2013 &    NaCo-$L\!'$       &  $305\pm13$         &  $-546\pm13$        &  $626\pm13$        &  $150.8\pm1.3$      & $9.7 \pm0.6$          &   $-0.58\pm0.10$   & $27.10\pm0.03$   & 1              \\ 
27-06-2013 &    NaCo-$L\!'$       &  $291\pm8$          &  $-525\pm8$         &  $600\pm11$        &  $151.0\pm1.2$      & $9.2 \pm0.8 $         &   $-0.65\pm0.10$   & $27.10\pm0.04$   & 1              \\ 
\noalign{\smallskip}\hline\noalign{\smallskip} 
10-12-2013 &    GPI-$K_1$         &  $301\pm5$          &  $-541\pm5$         &  $619\pm5$         &  $150.9\pm0.5$      &  $12.1\pm0.5$       &  $-0.10\pm0.13$    & $14.166\pm0.007$   & 2,3 \\  
11-12-2013 &    GPI-$H$           &  $306\pm11$         &  $-537\pm11$        &  $618\pm11$        &  $150.3\pm1.1$      &  $13.1\pm0.9$       &  $-0.10\pm0.13$    & $14.166\pm0.007$   & 2,3 \\  
13-05-2014 &    GPI-$K_1$         &  $307\pm8$          &  $-536\pm8$         &  $618\pm8$         &  $150.2\pm0.7$      &  -       &  $-0.10\pm0.13$    & $14.166\pm0.007$   & 3 \\  
06-04-2015 &    GPI-$K_1$         &  $322\pm7$          &  $-532\pm7$         &  $622\pm7$         &  $148.8\pm0.6$      &  -       &  $-0.10\pm0.13$    & $14.166\pm0.007$   & 3 \\  
08-04-2015 &    GPI-$K_1$         &  $320\pm4$          &  $-533\pm4$         &  $622\pm4$         &  $149.0\pm0.4$      &  $12.2\pm0.2$       &  $-0.10\pm0.13$    & $14.166\pm0.007$   & 3,4 \\  
29-02-2016 &    GPI-$H$           &  $330\pm5$          &  $-525\pm5$         &  $621\pm5$         &  $147.8\pm0.5$      &  $13.7\pm0.2$       &  $-0.10\pm0.13$    & $14.166\pm0.007$   & 3,4 \\  
06-03-2016 &    GPI-$H$           &  $336\pm3$          &  $-521\pm3$         &  $620\pm5$         &  $147.2\pm0.5$      &  -       &  $-0.10\pm0.13$    & $12.166\pm0.007$   & 3 \\  
\noalign{\smallskip}\noalign{\smallskip}\hline\noalign{\smallskip} 
03-02-2015 &    IRDIS-$K_1$       &  $322\pm4$          &  $-532\pm4$         &  $622\pm4$         &  $148.8\pm0.4$    &  $12.2\pm0.1$       &  $1.75\pm0.10$    & $12.26\pm0.01$   & 5\\  
03-02-2015 &    IRDIS-$K_2$       &  $319\pm5$          &  $-532\pm5$         &  $620\pm5$         &  $149.0\pm0.5$    &  $11.8\pm0.2$       &  $1.75\pm0.10$    & $12.26\pm0.01$   & 5\\  
05-05-2015 &    IRDIS-$K_1$       &  $324\pm7$          &  $-531\pm7$         &  $622\pm7$         &  $148.6\pm0.6$     &  $12.4\pm0.3$        &  $1.75\pm0.10$    & $12.26\pm0.01$   & 5\\  
05-05-2015 &    IRDIS-$K_2$       &  $322\pm8$          &  $-530\pm8$         &  $620\pm8$         &  $148.7\pm0.6$     &  $12.0\pm0.3$        &  $1.75\pm0.10$    & $12.26\pm0.01$   & 5\\  
18-01-2016 &    IRDIS-$K_1$       &  $327\pm8$          &  $-531\pm8$         &  $624\pm8$         &  $148.4\pm0.7$     &  $12.2\pm0.4$      &  $1.74\pm0.10$    & $12.26\pm0.01$   & 5\\  
18-01-2016 &    IRDIS-$K_2$       &  $326\pm10$         &  $-534\pm10$        &  $626\pm10$        &  $148.6\pm0.9$     &  $11.9\pm0.4$        &  $1.74\pm0.10$    & $12.26\pm0.01$   & 5\\  
31-05-2016 &    IRDIS-$K_1$       &  $334\pm3$          &  $-525\pm3$         &  $622\pm3$         &  $147.5\pm0.3$    &  $12.3\pm0.2$       &  $1.72\pm0.10$    & $12.26\pm0.01$   & 5\\  
31-05-2016 &    IRDIS-$K_2$       &  $332\pm4$          &  $-523\pm4$         &  $620\pm4$         &  $147.6\pm0.4$    &  $11.8\pm0.2$       &  $1.72\pm0.10$    & $12.26\pm0.01$   & 5\\  
10-05-2017 &    IRDIS-$K_1$       &  $343\pm3$          &  $-521\pm3$         &  $624\pm3$         &  $146.6\pm0.3$    &  $12.3\pm0.2$       &  $1.78\pm0.10$    & $12.26\pm0.01$   & 5\\
10-05-2017 &    IRDIS-$K_2$       &  $343\pm4$          &  $-524\pm4$         &  $626\pm4$         &  $146.8\pm0.4$    &  $12.3\pm0.2$       &  $1.78\pm0.10$    & $12.26\pm0.01$   & 5\\
\hline\noalign{\smallskip}
\end{tabular}
\tablefoot{References: (1) Astrometric and
  photometric results from \cite{rameau2013a} with the True North
  recalibrated on the 47 Tuc SHINE reference field ($+0.2\pm0.1\,\degr$) and the proper motion and precession correction applied for the parallactic angle calculation ($-0.15\pm0.05\degr$). See
  Appendix~A. (2, 3) Astrometric results processed by
  \cite{rameau2016} and photometric results from \cite{galicher2014},
  respectively.  (4) Photometric results reported by
  \cite{derosa2016}. (5) This work. Plate scale and True North from \cite{maire2016} (considering that \cite{maire2016} report the IRDIS True North correction and not the True North). }
\end{table*}

\begin{table}[t]
\caption{IFS Relative astrometry and photometry of HD\,95086 A and b}             % title of Table
\label{tab:ifsresults}
\centering
\begin{tabular}{lllll}     % 7 columns
\hline\hline\noalign{\smallskip}
Ins-Filter    & $\Delta\lambda$ & Sep.      &   PA                &  Contrast           \\
              & ($\mu$m)     &  (mas)       &   ($\degr$)             & (mag)               \\ 
\noalign{\smallskip}\hline\noalign{\smallskip} 
$J_{\rm{IFS}}$  & 1.20--1.32    &  $627\pm6$   & $147.9\pm0.5$      & $14.5_{-0.4}^{+0.5}$              \\ 
$H_{\rm{IFS}}$  & 1.60--1.65    &  $626\pm8$   & $147.8\pm0.7$      & $13.7\pm0.3$              \\ 
\noalign{\smallskip}\hline\noalign{\smallskip}
\end{tabular}
\end{table}

In the full IFS field of view (hereafter FoV;
$1.77\,''\times1.77\,''$), excluding the star, the unique point-source detected is the
planet HD\,95086\,b located at a separation of $\sim620$~mas and
position angle of $\sim148.0\degr$ from HD\,95086. The IFS
photometry and astrometry is discussed in Sec.\,\ref{sect:astro} and
\,\ref{sect:sed}. In the final IRDIS image (combining all rotated FoVs of
$11\,''\times11\,''$), together with HD\,95086\,b, 10 companion
candidates (ccs) are detected as shown in Fig.~\ref{fig:image} for
the May 31st 2016 epoch. All candidates are identified in the \textbf{five}
SHINE epochs with S/N\,$\ge 10$ (except cc-4 with a
S/N of $\sim5$). The cc relative astrometry and photometry with error bars
is derived using a standard fake planet injection with a $\chi^2$
minimization of the residuals within a segment of
$1\times$\,\textit{FWHM} extension in radius and
$3\times$\,\textit{FWHM} in azimuth \citep{chauvin2012}.  cc-8 corresponds to the bright
background star identified by \cite{rameau2013a} that served as a
sanity-check for the inter-astrometric calibration done between each
NaCo observation to confirm the discovery of HD\,95086\,b. None of the candidates detected correspond to the bright and faint sources resolved by ALMA deep 1.3 mm observations of \cite{su2017} and located near the edge of the \textit{cold} outer belt as seen in Fig.\,1. Same conclusions are drawn when going back to the NaCo $L\,'$ deep ADI observations of \cite{rameau2013a,rameau2013c}.

To control
and identify the presence of systematic biases related to the
platescale and True North correction in our SPHERE reduction, we measured the relative motion of all ccs (except cc-4) detected with IRDIS. This
aspect is critical as uncertainties on the platescale and True
North can significantly affect the determination of the relative
position of HD\,95086\,b, although closer than all ccs. A typical True
North systematic bias between two epochs of $0.2\,\degr$ will convert
into a systematic of 0.2~pixels (2.5\,mas) at HD\,95086\,b's
location ($\sim620~$mas), i.e the typical uncertainty we aim at to
monitor the planet's orbital motion given the good
signal-to-noise (S/N$\sim10$) achieved with IRDIS. Such a bias
scales into a 1.7~pixels shift for ccs at the edge of the IRDIS
FoV which would be easily spotted. In Fig.~\ref{fig:ppmtest}, we
report the differences between the relative astrometric positions of
all ccs between two SPHERE epochs: between February 3rd, 2015 and May
5th, 2015 (\textit{Top}) and between February 3rd, 2015, and May 31st,
2016 (\textit{Bottom}). The expected variation for stationnary background
objects is also reported with uncertainties based on the primary
parallactic and proper motions. Except HD\,95086\,b, all companion
candidates are clearly identified as background contaminants. Despite
the modest parallactic and proper motions of HD\,95086, we can see
that the companionship confirmation can be obtained with a good level
of confidence ($\ge 3 \sigma$) in 3 months owing to the
astrometric performances of SPHERE. We can also see that the
difference of relative positions of the ccs remain very compact
(within $\le4$~mas) conforting the good platescale and True North
astrometric calibration and the corresponding uncertainties (typically
$0.1\degr$ on the True North for each individual epoch). This verification was applied at each epoch to control systematic biases related to the platescale and True North correction in our SPHERE
reduction. As shown in Fig.~\ref{fig:ppmtest}, small systematic effects are still perceptible but within our error bars, varying from one epoch to another and likely imputable to the non-simultaneity of the science and astrometric calibration observations or to residual errors related to atmospheric and instrumental limitations.

%%%%%%%%%%%%%%%%%%%%%%%%%%%%%%%%%%%%%%%%%%%%%%%%%%%%%%%%%%%%%%%%%%%%%%
\section{Orbital properties of HD\,95086\,b}
\label{sect:astro}

\subsection{Relative astrometry}

Similarly to all ccs detected in the IRDIS FoV, the relative astrometry and
photometry of HD\,95086\,b was obtained in $K_1$ and $K_2$-bands using
a standard fake planet injection with a $\chi^2$ minimization of the
residuals within a segment of 1\,\textit{FWHM} extension in radius and
3\,\textit{FWHM} in azimuth (see Chauvin et al. 2012). The averaged
PSFs from the start/end sub-sequences were used for
injection. The results for each filter and epochs are reported in
Table~\ref{tab:results}. IRDIS results at all epochs were found
consistent between the IPAG-ADI pipeline using sADI and PCA and
the SHINE \textit{SpeCal} pipeline using TLOCI and PCA. They are reported together with
the NaCo relative astrometry of HD\,95086\,b obtained on January 12th,
2012, March 14th, 2013 and June 27th, 2013. The IFS astrometric results
are reported in Table~\ref{tab:ifsresults}. They are consistent with the IRDIS
results and are not considered in the following analysis given their
limited S/N and the fact they were obtained from images combining
several epochs.

\begin{figure}[t]
\hspace{0.2cm}
\includegraphics[width=\columnwidth]{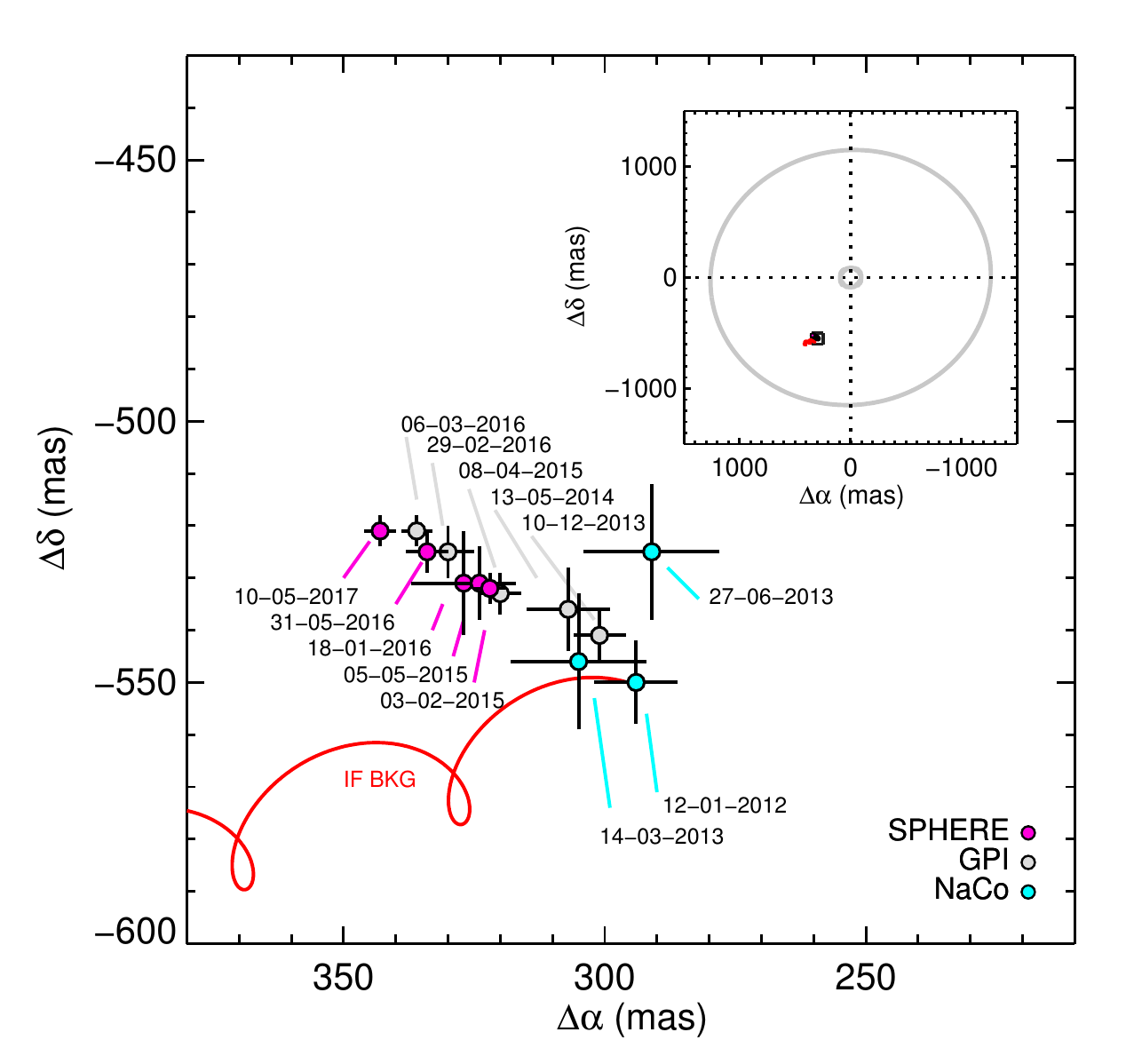}
\begin{centering}

\caption{Astrometric positions of HD\,95086\,b relative from A. The predictions for the HD\,95086\,b in case of
  a stationary background source are reported in red. SPHERE (pink) and GPI (grey) astrometric observations between 2013
  and 2017 are reported together with the NaCo observations
  (light-blue). The inset shows the planet location relative to the ones of the \textit{cold} outer belt and \textit{warm} inner belt from \cite{su2017}.}

\label{fig:orbit}
\end{centering}
\end{figure}

The NaCo astrometry was re-calibrated using contemporaneous
observations of the NaCo reference field we used since 2002 for
2M1207, $\beta$ Pictoris and HD\,95086 using the Orion stars TCC054,
58, 57, 34 and 26 \citep{mc1994,chauvin2012}
and the 47\,Tuc primary calibrator of SHINE \citep{maire2016}. These
observations were obtained in December 2014. We find a True North
correction of $+0.2\pm0.1\,\degr$ to add to the NaCo astrometry to
enable the homogenous calibration of NaCo and IRDIS data on
47\,Tuc. Finally, the NaCo astrometry published by \cite{rameau2013c} 
was in addition corrected from a bias of $-0.15\pm.05\,\degr$ related to the non consideration 
of the proper motion and precession correction in the equitorial coordinates of 
HD\,95086 for the parallactic angle calculation. 
GPI measurements published by \cite{rameau2016} are
reported in Table~\ref{tab:results} for comparison and discussion
about both instrument astrometric performances. They actually refer to
\cite{derosa2016} for the platescale and True North solutions
determined by continually observing a set of astrometric calibrators
with well-determined orbital solutions or contemporaneous NIRC2
measurements. As we cannot exclude at this stage the presence of
systematic biases between SPHERE and GPI as our reference calibrators
are not the same, we did not consider the GPI results in the following
orbital fitting analysis. NaCo, IRDIS and GPI astrometric results are shown together in
Fig.~\ref{fig:orbit}. Expected variation of offset positions in case
of a background object is reported in \textit{red} and unambiguously
confirms that HD\,95086\,b is co-moving. The NaCo (re-calibrated) and
IRDIS measurements independently confirm that the orbital motion is
now resolved with a high level of confidence as reported by
\cite{rameau2016}. Although GPI and SPHERE observations were not taken
at the same epochs, we can try to link the closest measurements in
time. Some small variations can be seen between almost contemporaneous
GPI measurements or IRDIS and GPI measurements, but they all remain
within the $1\sigma$ error bars. They independently demonstrate the
unprecented astrometric accuracy (3~mas in the best cases here) and the 
calibration procedure achieved by the new generation of planet imagers.

%%%%%%%%%%%%%%
\subsection{Orbital fitting}

\begin{figure*}[t]
\hspace{0.2cm}
\includegraphics[width=\textwidth]{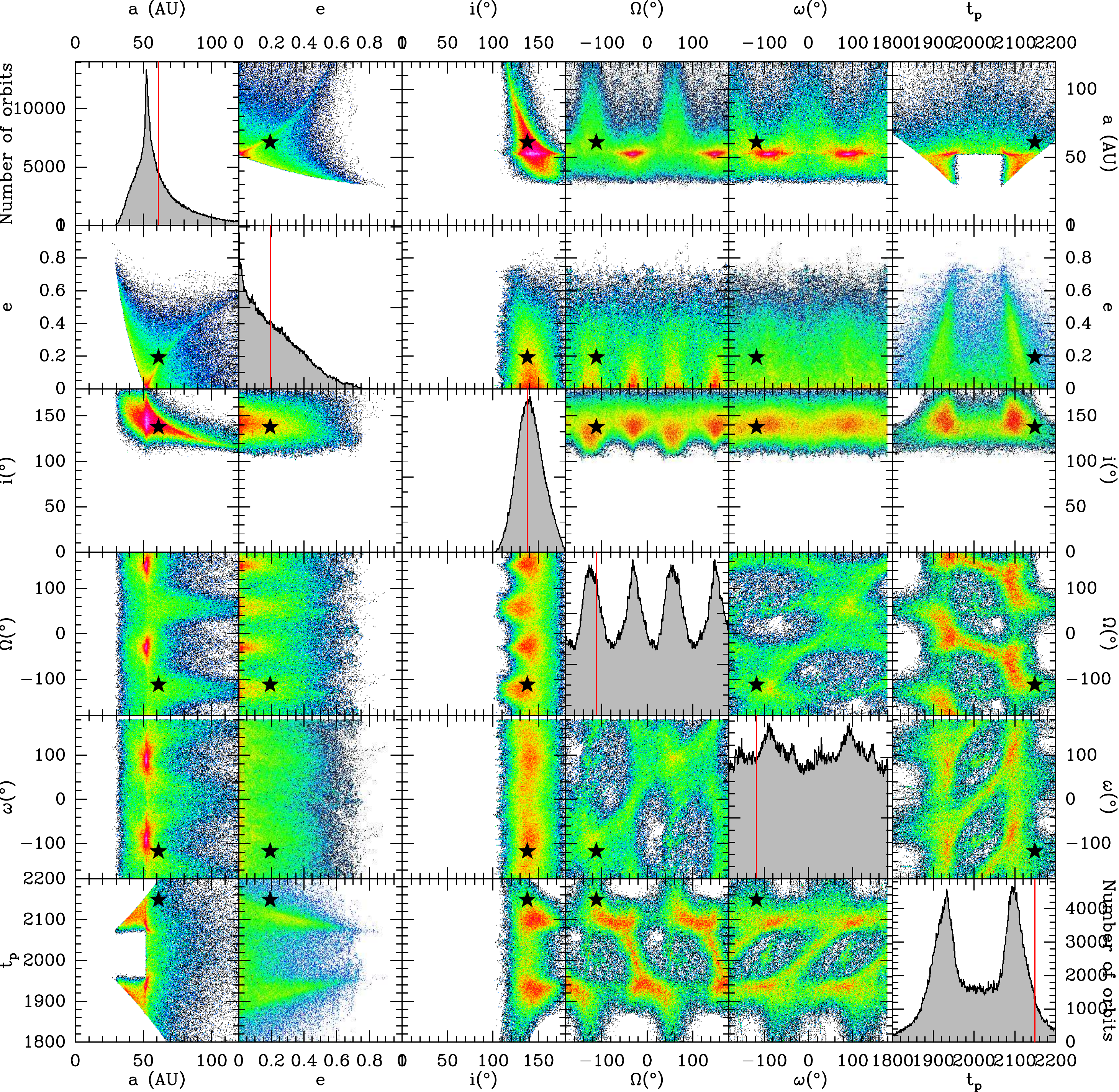}
\begin{centering}
\caption{Results of the MCMC fit of the NaCo and SPHERE combined astrometric data of
  HD\,95086\,b reported in terms of statistical distribution matrix of
  the orbital elements $a$, $e$, $i$, $\Omega$, $\omega$ and
  $t_p$. The \textit{red} line in the histograms and the \textit{black} star in the correlation plots indicate the position of the best LSLM
  $\chi_r^2$ model obtained for comparison.}
\label{fig:mcmc}
\end{centering}
\end{figure*}

As mentioned previously, we only considered the NaCo and SPHERE data
points to exclude any possible systematics in the orbital fitting
analysis. Following the method developed for $\beta$~Pictoris
\citep{chauvin2012}, we used a Markov-Chain Monte-Carlo (MCMC)
Bayesian analysis technique \citep{ford2005,ford2006}, which is well suited for observations covering a small part
of the whole orbit (for large orbital periods). This is the case for HD\,95086\,b 
as illustrated in the insert of
Fig.~\ref{fig:orbit}. We did not consider any prior information on the
inclination or longitude of ascending node to explore the full orbital parameter space of bound orbits. As described in Appendix\,A of \citep{chauvin2012}, we assume the prior distribution $p_0(\vec{x})$ to be uniform in $\vec{x}=(\log P, e, \cos i,\Omega+\omega,\omega-\Omega, t_p)$ and work on a modified parameter vector $\vec{u(x)}$ to avoid singularities in inclination and eccentricities and improve the convergence of the Markov chains. The results
of the MCMC analysis are reported in Fig.~\ref{fig:mcmc}, together with the results of a classical Least-squared linear method (LSLM) flagged by the \textit{red} line. It shows
the standard statistical distribution matrix of the orbital elements
$a$, $e$, $i$, $\Omega$, $\omega$ and $t_p$,
where $a$ stands for the semi-major axis, 
%(more adapted by definition than the semi-major
%axis $a$ for the representation of bound and unbound orbits),
$e$
for the eccentricity, $i$ for the inclination, $\Omega$ the longitude
of the ascending node (measured from North), $\omega$ the argument of
periastron and $t_p$ the time for periastron passage.

\begin{table}[t]
\caption[]{MCMC solutions for the orbital parameters of HD\,95086\,b: semi-major axis ($a$), period
  ($P$), eccentricity ($e$), inclination ($i$), longitude of ascending
  node ($\Omega$), argument of periastron ($\omega$) and time of periastron passage ($t_p$).}
\label{mle}
\centering                          % used for centering table
\begin{tabular}{ccc}
\noalign{\smallskip}\hline\noalign{\smallskip}
Parameter & Unit & MCMC solutions \\
\noalign{\smallskip}\hline\noalign{\smallskip}
\noalign{\smallskip}$a\,$ &(AU) & $52.0_{-24.3}^{+12.8}$ \\
\noalign{\smallskip}$P\,$ &(yr) & $288.6_{-176.5}^{+11.5}$ \\
\noalign{\smallskip}$e$ & & $0.2_{-0.2}^{+0.3}$ \\
\noalign{\smallskip}$i$ & ($\degr$) & $140.7_{-13.3}^{+14.8}$ \\
\noalign{\smallskip}$\Omega$ & ($\degr$) &  Peaks at $-31.4\pm180\degr$\\
                  &  &  and $-118.9\pm180\degr$ \\
\noalign{\smallskip}$\omega$ & ($\degr$) & $\sim$Flat distribution \\
\noalign{\smallskip}$t_p\,$ & (yr JD) & Peaks at 1933.1 and 2093.0 AD\\ 
\noalign{\smallskip}\hline
\end{tabular}
\end{table}

\begin{figure}[t]
\hspace{0.2cm}
\includegraphics[width=\columnwidth]{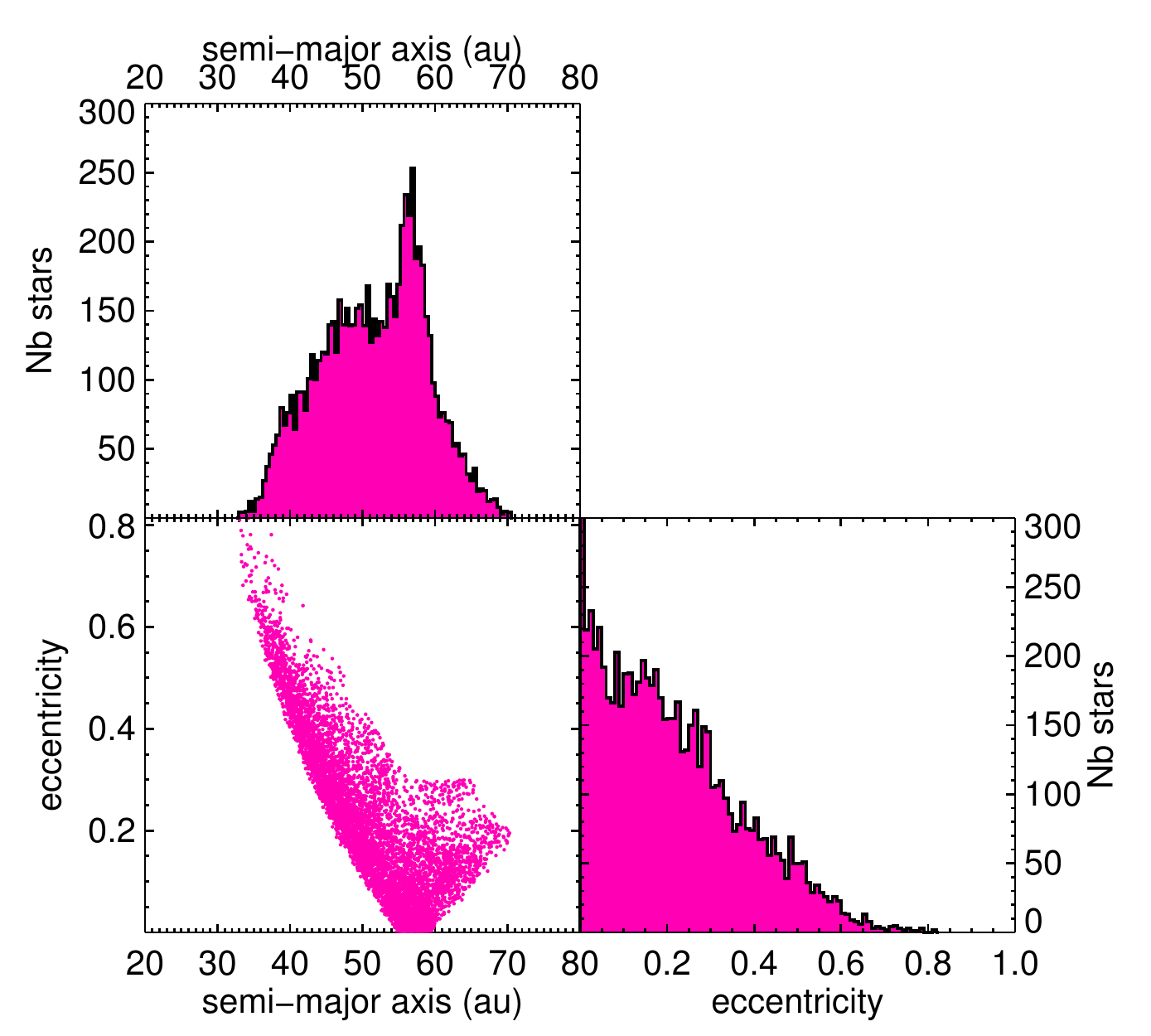}
\begin{centering}
\caption{Orbital solution of HD\,95086\,b assuming a coplanar configuration with the outer belt inclination ($i_{\rm{disk}}=130\pm3\degr$, $\Omega=98\pm 3\degr$, \citealt{su2017}), an inner extent of the the chaotic zone larger than 10~au, and an outer extent of the chaotic zone smaller than 106\,au, i.e. the inner boundary of the outer belt resolved by ALMA 1.3 mm observations.}
\label{fig:lazz}
\end{centering}
\end{figure}

The results of our MCMC fit indicate orbital
distributions that peak with a 68\% confidence interval at
$52.0_{-24.3}^{+12.8}$~au for the semi-major axis,
$140.7_{-13.3}^{+14.8}\degr$ for the inclination, eccentricities that
fall within $e=0.2_{-0.2}^{+0.3}$. The longitude of ascending node
shows several peaks at $\Omega -31.4\pm 180\degr$, $\Omega=-118.9\pm 180\degr$, 
but is not very strongly constrained, as well as the argument of periastron $\omega$.
Periastron passages occur between 1800 and 2200 AD with two peaks at 1933.1 and 2093.0~AD. 
The latter parameters are badly constrained basically because the set of acceptable 
orbital solutions is compatible with circular orbits and also with $i=180\degr$. 
Nonetheless, the inclination distribution cleary favors retrograde orbits ($i>90\degr$), 
which is compatible with the observed
clockwise orbital motion resolved with NaCo, GPI and SPHERE. 
The rather loose error bars directly result from the fact that we are
fitting only a very small portion of the whole orbit corresponding to
an almost linear orbital variation described by four parameters (position and
velocity in the projected celestial plane) without any constraints in
the third spatial dimension (the line of sight). The fact that favored
solutions of periastron passages fall at the current epoch likely
result from the limited number of observational constraints biasing
the orbital fitting process.  

Coplanar solutions with the outer belt plane resolved by ALMA
($i_{\rm{disk}}=130\pm3\degr$, $\Omega=97\pm 3\degr$, \citealt{su2017}) are
compatible with the posterior distribution of orbits, although not falling at the current peak of the distribution. We defined them as coplanar when the mutual inclinations between the planet orbital plane and the outer belt plane lie within $\pm5\degr$. The mutual inclination ($i_{\rm{tilt}}$) is given by 

\begin{equation}
\label{eq:mut}
\cos(i_{\rm{tilt}})=\cos(i)\cos(i_{\rm{disk}})+\sin(i)\sin(i_{\rm{disk}})\cos(\Omega-\rm{PA}_{\rm{disk}}) 
\end{equation}

The correponding coplanar solutions are shown in Fig.\,B.1 of Appendix\,\ref{App:A}. They favor orbit with lower eccentricities and smaller semi-major axis. If we exclude in addition all coplanar solutions for which an outer extent of the chaotic zone created by HD\,95086\,b lies below 10\,au or beyond 106\,au (following \citealt{lazzoni2017} prescriptions), we restrain even more the ($a$, $e$) parameter space compatible with our NaCo and SPHERE observations, and the the inner edge of the \textit{cold} outer belt as seen in Fig.\,\ref{fig:lazz}). 

%The
%distribution of the longitude of ascending node becomes also less
%constrained. 

Globally, our orbital fitting analysis is consistent with the results
of \cite{rameau2016} (see their Table~1 and Fig.~3 considering a correction factor of 0.92 in sma and 0.89 in period given the new \textit{Gaia} distance), although less
constraining regarding the values within the 68\% confidence
interval. Both favor retrograde low- to moderate-eccentricity solutions with
semi-major axis peaking at $50-55$~au that are compatible within the
error bars with a coplanar configuration with the disk inclination. The small differences between \cite{rameau2016} and this work may arise from the non complete overlap in terms of values and uncertainties between SPHERE and GPI in addition to the new \textit{Gaia} distance revision. To verify the possible discrepancies due to the use of different MCMC orbital fitting approaches in this work \citep{beust2016} and by \cite{rameau2016} refering to the Orbits For Impatients Tools (OFTI) developed by \citep{blunt2017}, we ran as a check our MCMC analysis on the same set of NaCo and GPI data used by \cite{rameau2016}. The results are presented in Fig.\,C.1 of Appendix\,\ref{App:B}, and can be compared with Fig.\,3 of \cite{rameau2016}. There is a very good match between the distributions of parameters found by both tools. We see some minor variations for the eccentricity distribution possibly linked to the difference of prior, and for the
  distribution of time of periastron passage probably related to the non-renormalization of the distribution by \cite{rameau2016} over a given range of orbital period.

%%%%%%%%%%%%%%%%%%%%%%%%%%%%%%%%%%%%%%%%%%%%%%%%%%%%%%%%%%%%%%%%%%%%%%
\section{Spectral energy distribution of HD\,95086\,b}
\label{sect:sed}

%
%-------------------------------------------------------------------
\subsection{Conversion to fluxes}
\label{sec:convf}

\begin{figure*}[t]
\begin{center}
\begin{tabular}{cc}
\includegraphics[width=8cm]{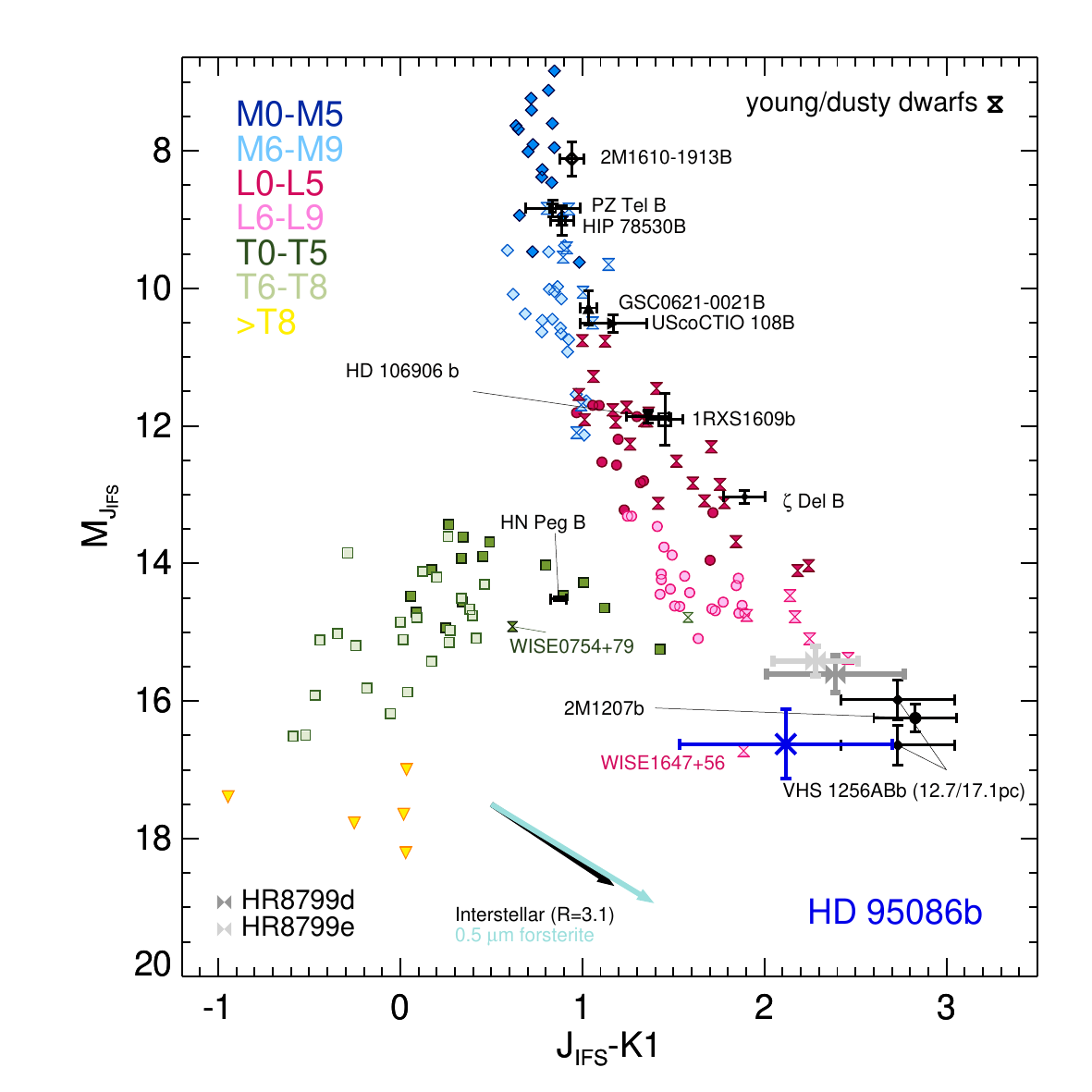} & 
\includegraphics[width=8cm]{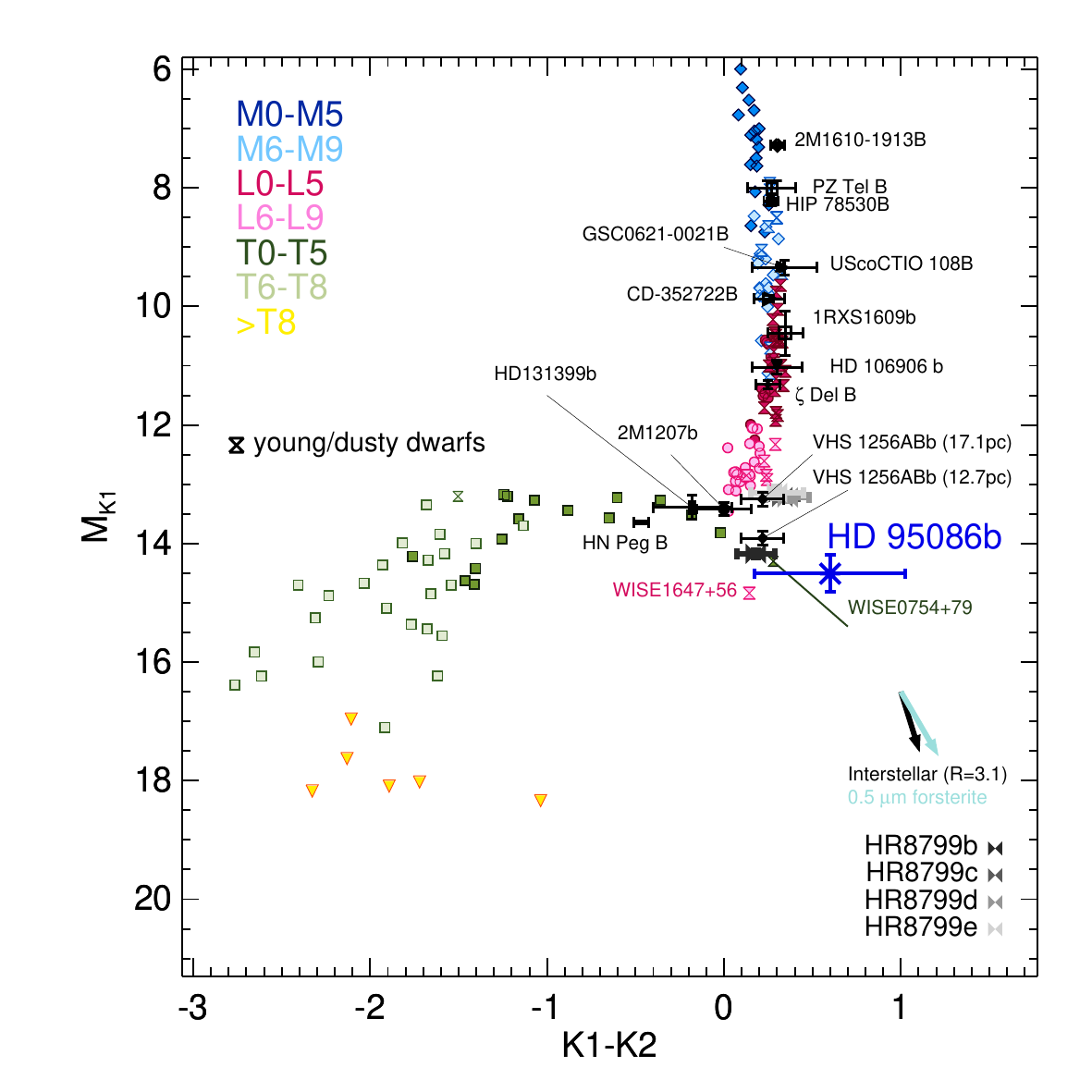} \\
\end{tabular}
\caption{\textit{Left:} Color-magnitude diagram considering the SPHERE/IRDIS $K1$ photometry, and the $J_{\rm{IFS}}$ photometry from 1.2 to 1.32 $\mu$m  as it can be extracted from the SPHERE/IFS datacubes. \textit{Right:} Color-magnitude diagram considering the SPHERE/IRDIS $K1$ and $K1$ photometry. In both cases, Measurements of M, L, T field dwarfs, and of
young companions and red dwarfs are reported. Reddening vector computed for the synthetic interstellar extinction curve for a 0.5\,mag $K$ band extinction and reddening parametrer of $R_v = 3.1$ is overlaid (\textit{black arrow}). We also report the reddening vector from the extinction curve of forsterite dust grains with size distribution
centered on radii $r = 0.5\,\mu$m and for a 0.15\,mag $K$ band extinction (\textit{light blue arrow}). As proposed by \citep{2014MNRAS.439..372M, 2016A&A...587A..58B}, both could explain the deviation of the colors of young and dusty L6-L8 dwrafs, and of the HR\,8799 planets with respect to the sequence of field dwarfs. The photometry of HD~95086\,b shown by the \textit{blue cross} with uncertainties in both diagrams is characteristic of young L/T transition objects.}

\label{Fig:CMD}
\end{center}
\end{figure*}

%\subsection{Comparison to reference objects}
%\label{sec:refobj}
We used a spectral template from the BT-NEXTGEN library
\citep{2012RSPTA.370.2765A} degraded to R$\sim$30 (the smallest
resolution of our new observations) and adjusted it onto the TYCHO B
and $V$, 2MASS $J$,$H$, $K_{s}$, and WISE $W1$ and $W2$ photometry
\citep{2000A&A...355L..27H, 2003tmc..book.....C,
  2012yCat.2311....0C}. The star photometry is best represented by a
model with $T_{\rm{eff}}$=7600 K, and log(g)=4.0. We assumed a metallicity
of [M/H]=0 supported by the abundance analysis of a few LCC
  members by \cite{vianaalmeida2009}. Those parameters are close to those found by
\cite{2013ApJ...775L..51M} and considered by \cite{derosa2016}.  The
fit is shown in Fig.~\ref{Fig:SEDstar} of Appendix~\ref{app:convflux}. We used this spectrum to compute the average stellar flux at the wavelengths of 
the $J_{\rm{IFS}}$ and $H_{\rm{IFS}}$ SPHERE IFS passbands (Table\,\ref{tab:ifsresults}), 
and through the IRDIS $K1$ and $K2$, and GPI $H$ and $K1$ IFS passbands. In addition to the SPHERE IFS relative photometry of Table\,\ref{tab:ifsresults}, we took the
weighted mean of the GPI $H$ and $K1$, SPHERE $K1$, and SPHERE $K2$ contrast values
reported in Table\,\ref{tab:results} to compute the photometry of HD95086\,b into
the corresponding passbands. SPHERE and GPI passbands do overlap at $H$ and $K$-band, but do not cover the same spectral range. They were therefore independently considered for the spectral analysis. Finally, we considered the remaining flux values from low resolution GPI $K1$ spectrum and NaCo $L\!'$ photometry of Table\,1 of \cite{derosa2016}.

%
%-------------------------------------------------------------------
\subsection{Color-magnitude diagrams}
\label{sec:refobj}

We report in Fig. \ref{Fig:CMD} the location of HD95086\,b in $J$-band
and $K$-band based color-magnitude diagrams (CMD).  Details on the
diagrams are given in \cite{2016A&A...593A.119M} and \cite{samland2017}. We used here the most recent parallaxes of the young objects
from \cite{2016ApJ...833...96L}, and added additional companions
\citep{2015ApJ...804...96G, 2016ApJ...818L..12S, 2014MNRAS.445.3694D}
at the L/T transition, and/or members of Sco-Cen \citep[][and ref.
  therein]{2015ApJ...802...61L, 2014ApJ...780L...4B,
  2016Sci...353..673W}. HD~95086\,b falls at the L/T transition but is
underluminous compared to the field dwarfs. The underluminosity is
characteristic of young L/T objects \citep[see Fig~25 of][and
  ref. therein]{2016ApJ...833...96L}. Its placement is marginally
consistent in both diagrams with VHS J125601.92-125723.9 ABb
\citep[age $\leq$ 320 Myr, L7, ][]{2015ApJ...804...96G} if the
distance to the system is 12.7 pc \citep[][]{2016ApJ...818L..12S}. The
planet has a photometry compatible with the one of the peculiar L9
dwarf WISE J164715.57+563208.3 (Appendix C; assuming that
the parallax reported in \cite{2011ApJS..197...19K} is robust). The
latter has recently been proposed to be a 4-5 M$\mathrm{_{Jup}}$
free-floating member of the Argus association, i.e. an object with a
mass in the same range as HD~95086\,b. To conclude, the planet falls in
the K1-K2 diagram close to HR~8799b and WISE0754+49, two objects with
early T spectral types and red colors. HR~8799b has a mass close to
the one of HD~95086\,b (estimated at 4-5 M$\mathrm{_{Jup}}$). We overlaid in the diagrams the
reddenning vectors caused by interstellar extinction
\citep{2003ARA&A..41..241D}
and by 0.5 $\mu$m forsterite grains which are proposed to explain the
red colors of the dusty and/or variable L dwarfs
\citep{2014MNRAS.439..372M, 2016A&A...587A..58B, 2016ApJ...830...96H,
  2016ApJ...829L..32L}. For the forsterite grains, we used the optical constants of \cite{scottduley1996}.
The young and/or dusty L/T object photometry
is shifted along those vectors \citep[see][]{2016A&A...587A..58B} with
respect to the sequence of field dwarfs, and so does HD~95086\,b.
%
%-------------------------------------------------------------------
\subsection{Empirical spectral comparison}

\begin{figure}[t]
\begin{center}
\includegraphics[width=\linewidth]{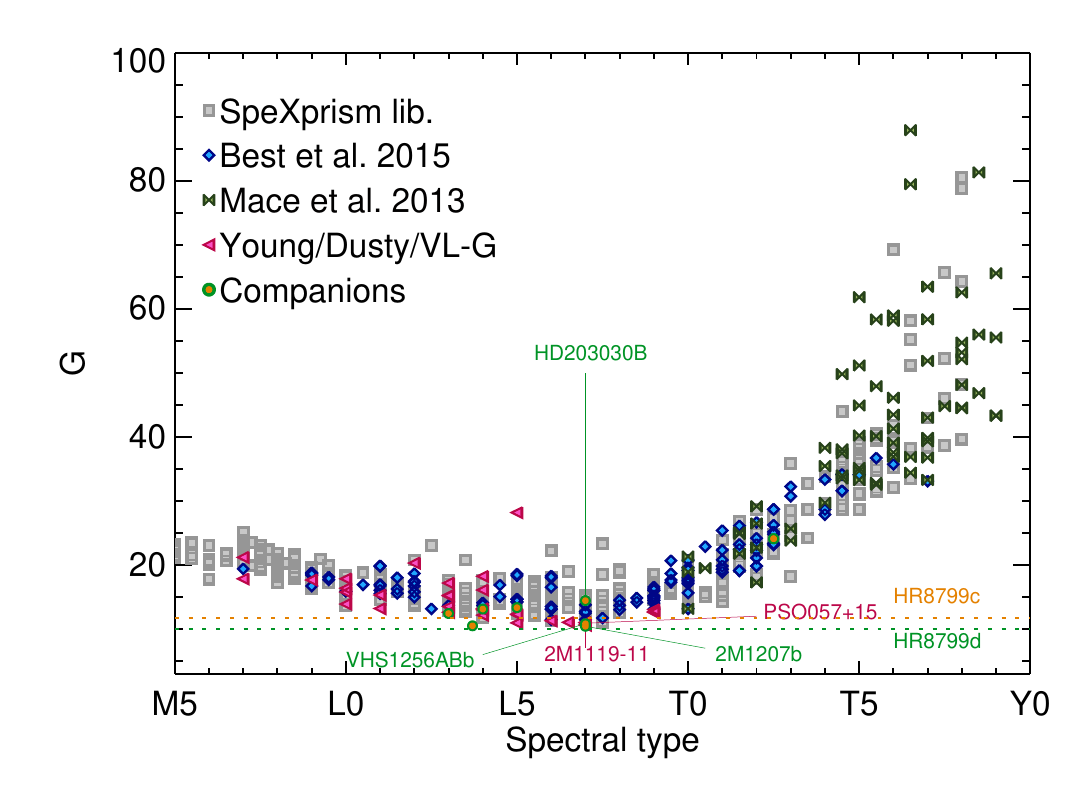}
\caption{Evolution of G with the spectral type for the different libraries of comparison objects considered.}
\label{Fig:G}
\end{center}
\end{figure}

We compared the $J$ to $K$-band spectrophotometry of HD\,95086\,b to the one of comparison
objects generated from low resolution ($R_\lambda\sim75-120$) spectra acquired
with the SpeX instrument. The spectra considered were taken from the
SpeXPrism library \citep[][532 objects]{2014ASInC..11....7B},
\cite{2015ApJ...814..118B} (122 objects), and
\cite{2013ApJS..205....6M} (72 objects). The first two libraries
include spectra of peculiar "dusty" L dwarfs. We also considered the
spectra of young M and L dwarfs of \cite{2013ApJ...772...79A} (17
spectra), and of red/dusty L dwarfs later than L4 taken from the
literature (see Appendix \ref{App:C}). To conclude, we included in
the analysis the spectra of the exoplanets HR8799c and d
\citep{2013ApJ...768...24O, 2015ApJ...803...31P, 2016A&A...587A..57Z},
and of the companions younger than the population of field objects and
spanning the L/T transition (age$\leq$400 Myr): VHS
J125601.92-125723.9 ABb, 2M1207b \citep{2010A&A...517A..76P}, HR8799d
and e \citep{2016A&A...587A..57Z}, HIP203030B (Bonnefoy et al.  in
prep), $\zeta$ Del B \citep{2014MNRAS.445.3694D}, 2M0219-39b
\citep{2015ApJ...806..254A}, 2M0122-24B \citep{2015ApJ...805L..10H},
and G196-3B \citep{1998Sci...282.1309R}.  

All spectra were smoothed to
a resolution of $R_\lambda\sim$66, corresponding to the resolution of the GPI
spectrum\footnote{The SPHERE and P1640 $JH$ spectra were not smoothed
  because their original resolution is lower than the one of
  GPI. Nonetheless, the spectrophotometry of HD~95086\,b in the $J$ and $H$
  band is lower than the SPHERE and P1640 data.}. We considered the
$G$ goodness-of-fit indicator defined in \cite{2008ApJ...678.1372C}
which accounts for the filter and spectral channel widths to compare
each of the $k$ template spectra to the $n$ spectrophotometric datapoints of
HD~ 95086\,b

\begin{figure}[t]
\begin{center}
\includegraphics[width=\linewidth]{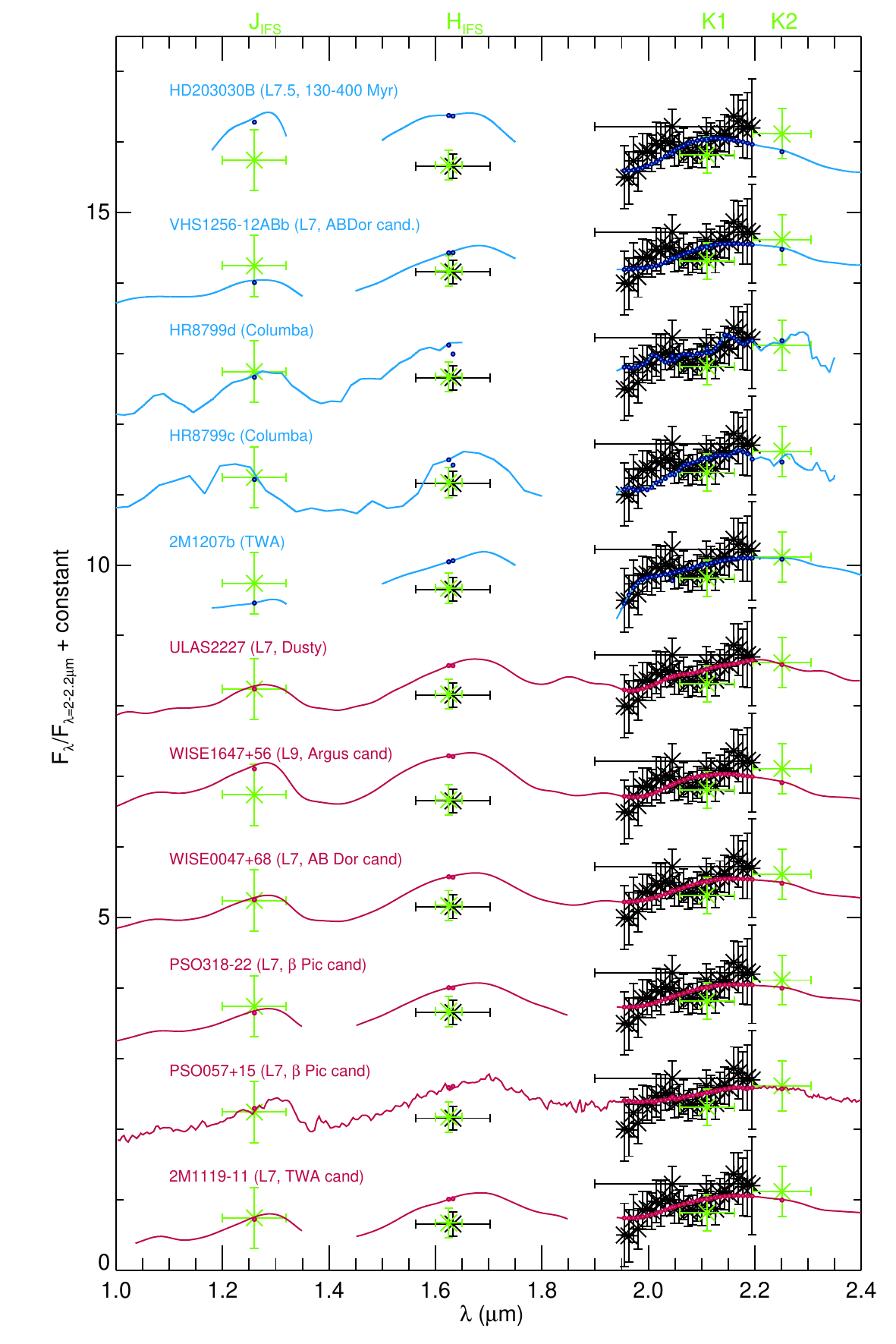}
\caption{Spectrophotometry of HD~95086\,b (SPHERE IFS $J_{IFS}$, $H_{IFS}$, and IRDIS $K1$ and $K2$ photometric datasets in \textit{green}, GPI $H$ and $K1$ photometric datasets in \textit{purple}, and GPI low-resolution $K1$ spectrum in \textit{black}) compared to near-infrared spectra of young and/or dusty L7-L7.5 dwarfs (dark red) and companions (blue) younger than the field. The L7 dwarfs candidate members of young moving groups are sorted by their supposed age.}
\label{Fig:visufit}
\end{center}
\end{figure}

\begin{figure}[t]
\begin{center}
\includegraphics[width=\linewidth]{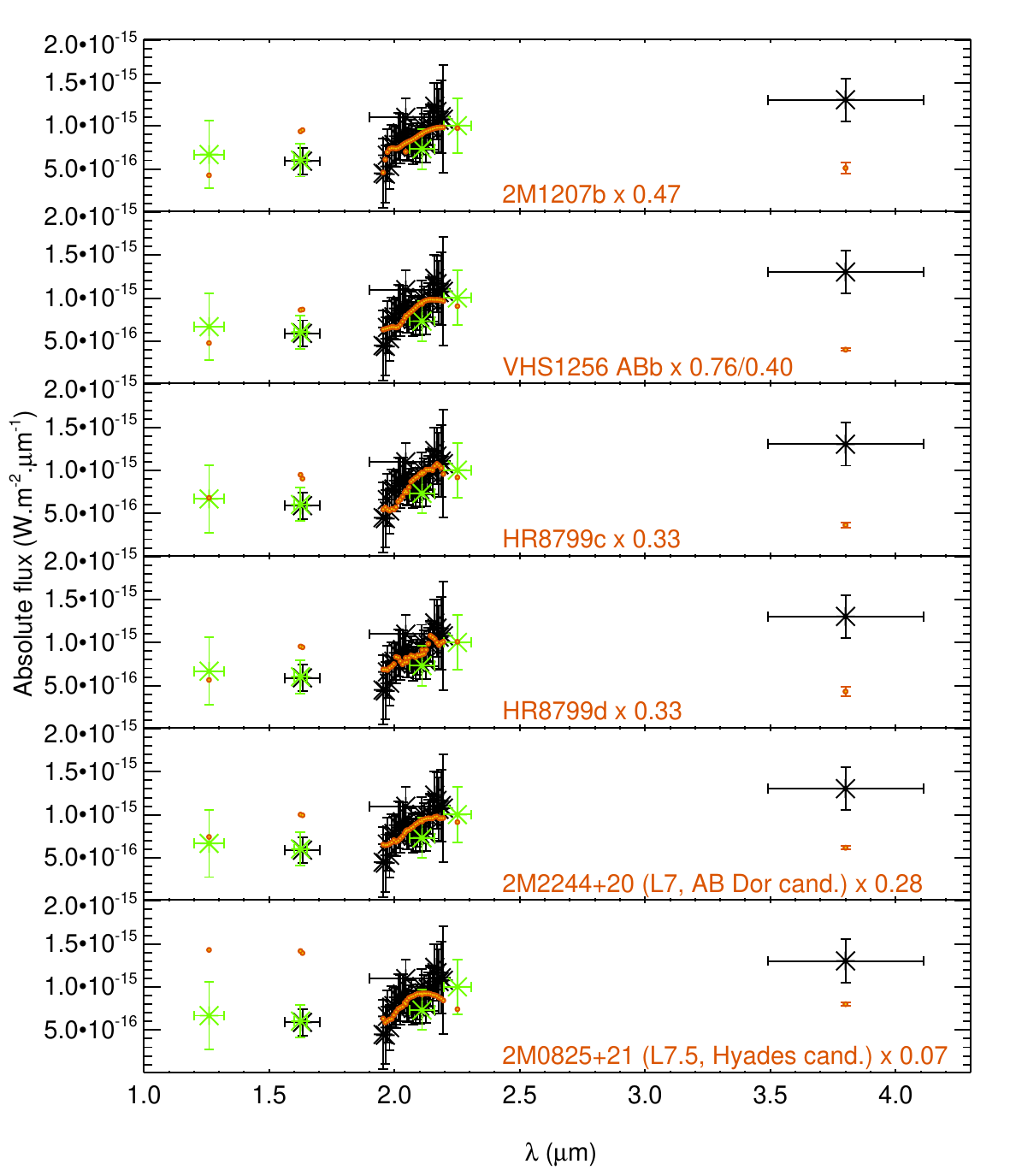}
\caption{The 1-4 $\mu$m spectral energy distribution of HD~95086\,b (SPHERE IFS $J_{IFS}$, $H_{IFS}$, and IRDIS $K1$ and $K2$ photometric datasets in \textit{green}, GPI $H$ and $K1$ photometric datasets in \textit{purple}, GPI low-resolution $K1$ spectrum in \textit{black}, and NaCo $L\!'$ photometric datset in \textit{red}.) compared to the one of young and/or dusty L7-L7.5 dwarfs and companions (dark red) younger than the field.}
\label{Fig:SED}
\end{center}
\end{figure}
\begin{equation}
\label{eq:gk}
G_{k}=\sum_{i=1}^{n} w_{i} \left ( \frac{f_{i} - \alpha_{k}\textsl{F}_{k,i}}{\sigma _{i}} \right )^{2},
\end{equation}  

where $f_{i}$ and $\sigma _{i}$ are the observed data point $i$ and
associated error, $w_{i}$ is the filter width of the corresponding
data point. $\textsl{F}_{k,i}$ is the value of the photometry in the
given filter passband $i$ or GPI IFS wavelength channel for the
template spectrum $k$. $\alpha_{k}$ is a multiplicative factor between
the planet spectrophotometry and the one of the template which
minimizes $G_{k}$ and is given by

\begin{equation}
\label{eq:alpha}
\alpha_{k} = \frac{\sum_{i=1}^{n}w_{i}f_{i}\textsl{F}_{k,i}/\sigma _{i}^{2}}{\sum_{i=1}^{n}w_{i}\textsl{F}_{k,i}/\sigma _{i}^{2}},
\end{equation}  
 
The GPI spectrum is affected by correlated noise \citep{2016ApJ...833..134G}. \cite{derosa2016} accounted for this noise in the fit

\begin{equation}
G_{k}=R_{k}^{T}C^{-1}R_{k} + \sum_{i=1}^{n} w_{i} \left ( \frac{f_{i} - \alpha_{k}\textsl{F}_{k,i}}{\sigma _{i}} \right )^{2},
\end{equation}  

where $C$ is the correlation matrix determined following
\cite{2016ApJ...833..134G} and $R_{k} = f - \alpha_{k}\textsl{F}_{k}$.
\cite{derosa2016} found $\alpha_{k}$ via a truncated-Newton
algorithm. We preferred to find the one minimizing $G_{k}$,
considering values of $\alpha_{k}$ plus or minus 100 times the one
found with the Eq. \ref{eq:alpha}. This exploration of $\alpha_{k}$ is
sufficient to keep the values of $G_{k}$ unchanged for all the
considered comparison objects.

\begin{figure*}[t]
\hspace{0.2cm}
\includegraphics[width=5.5cm]{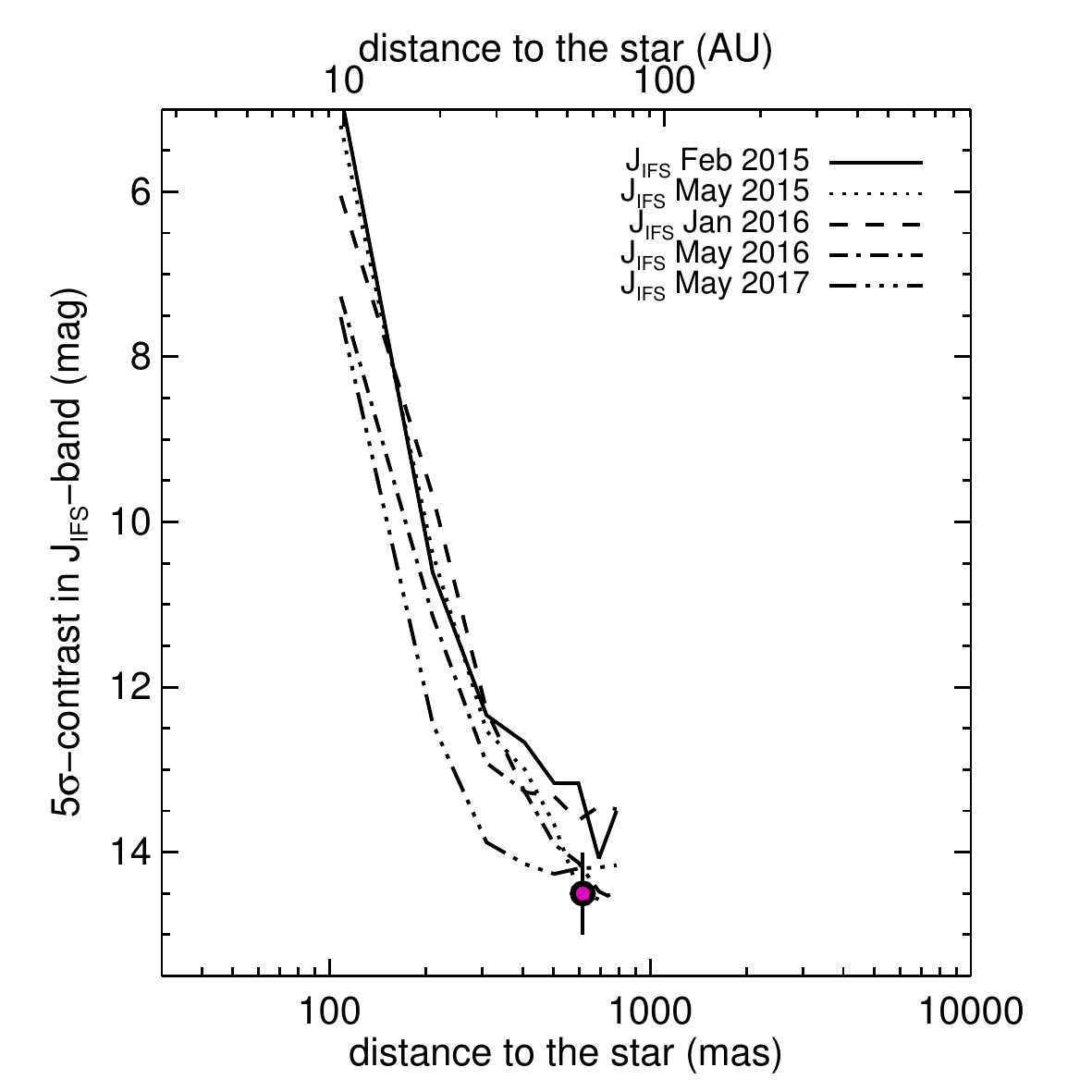}
\includegraphics[width=5.5cm]{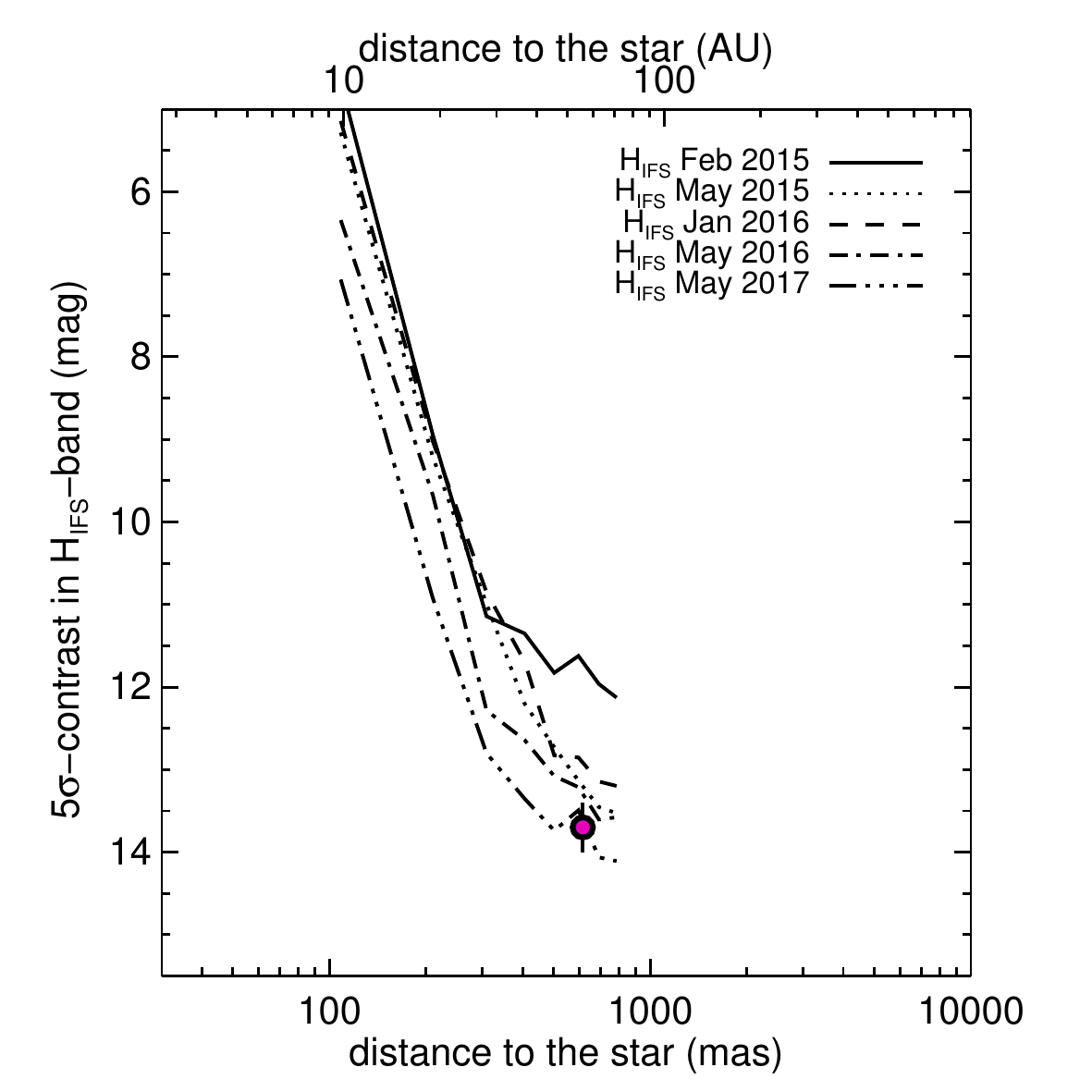}
\includegraphics[width=5.5cm]{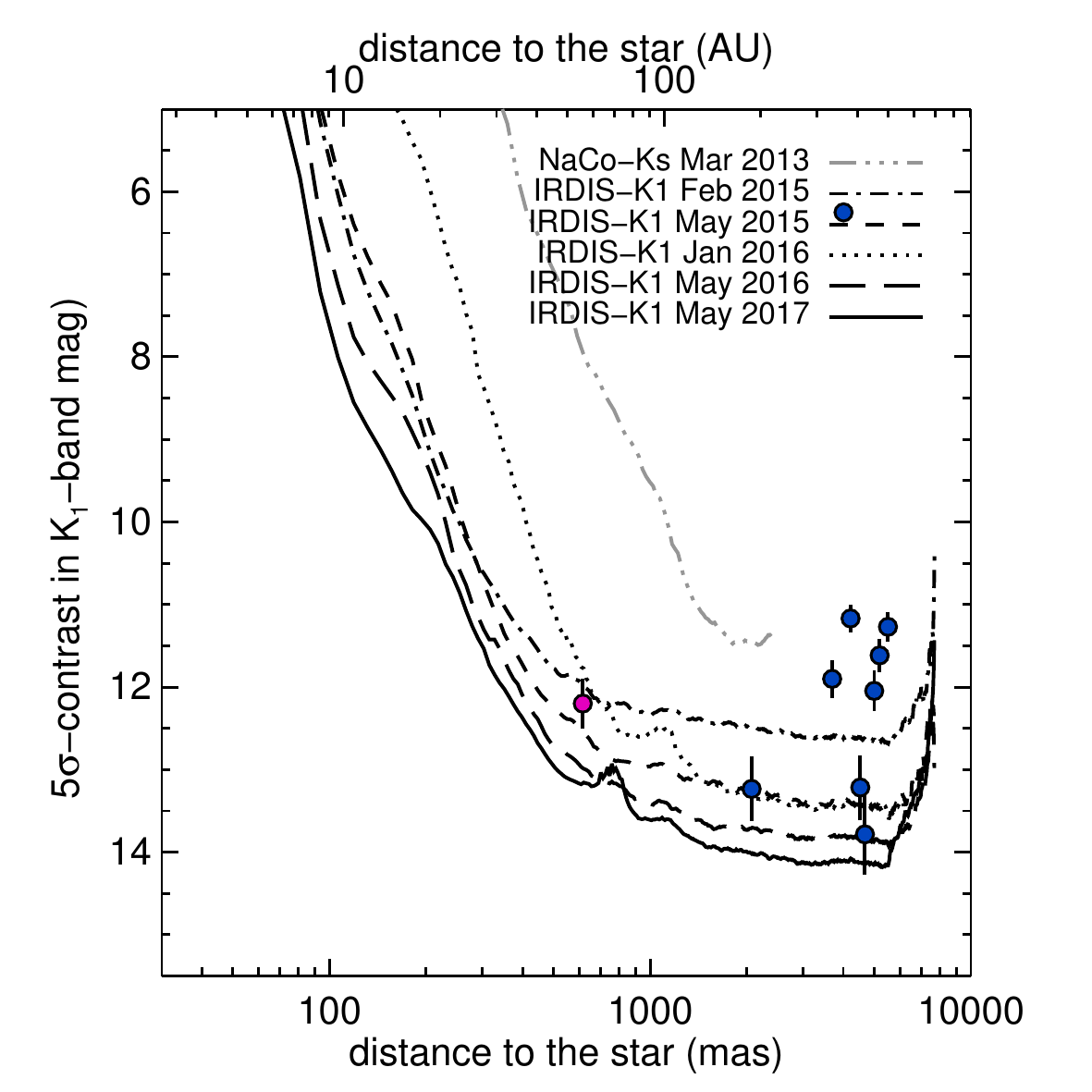}
\begin{centering}

\caption{\textit{Left:} IFS $J$-band detection limit at $5\sigma$
  based on the spectral PCA analysis using 150 eigen modes obtained in
  February 3rd, 2015, May 5th, 2015, January 18th, 2016, May 31st, 2016, and May 10th, 2017.
  Flux loss, coronagraphic transmission
  and small statistics effects have been corrected. The detection of
  HD\,95086\,b combining all epochs is reported in \textit{pink} with
  error bars. \textit{Middle:} Same as before in $H$-band for the
  IFS. The planet is also reported in \textit{pink} with error
  bars. \textit{Right:} Same but for the IRDIS $K1$-band detection limits reported together with all the point-sources detected in the field
  of view identified as background objects and HD\,95086\,b reported
  in \textit{pink} with error bars.}

%% \caption{\textit{Left:} HARPS detection limits calculated using the local power
%%   analysis method \citet{meunier2012}. \textit{Middle:} IRDIS $K1$-band full-combined image of
%%   HD\,95086 from May 31st 2016. All companion candidates have been
%%   anooted. HD\,95086\,b is well detected at a separation of
%%   $618\pm3$~mas and position angle of $147.5\pm0.2$~deg from
%%   HD\,95086. \textit{Right:} Zoom-in of IRDIS $K1$ and $K2$-band
%%   images of HD\,95086 and HD\,95086\,b observed in February 3rd 2015,
%%   May 5th 2015 and May 31st 2016.}
\label{fig:detlim}
\end{centering}
\end{figure*}
We show in Fig.~\ref{Fig:G} the evolution of $G$ with the spectral
type for the different libraries of templates considered. We find that
the planet photometry is best fitted by dusty and/or young L5-L7.5
dwarfs. Among the SpeXprism library, we retrieve a best fit for the L7
dwarf 2MASSWJ2244316+204343, a candidate member of the AB Dor moving
group (see Appendix \ref{App:C}). The dusty L/T transition object
WISEJ064205.58+410155.5 is the template from the
\cite{2013ApJS..205....6M} library which provides the best fit to the
planet spectrophotometry. This object is a possible member of the AB
Dor moving group \citep{2015ApJ...798...73G}. The best fits for the
\cite{2015ApJ...814..118B} library and among the dusty L dwarfs
considered (Appendix \ref{App:C}) are obtained for
PSO\_J057.2893+15.2433 and 2MASS J11193254-1137466, two L7 objects
which are candidate members of the $\beta$ Pictoris moving group and
TW Hydrae association, respectively. These objects have predicted
masses of 6.6-9.9 and 4.3-7.6 $\mathrm{M_{Jup}}$, respectively
\citep{2015ApJ...814..118B, 2016ApJ...821L..15K}. To conclude, we find
that the companions HR8799d, 2M1207b, and VHS J125601.92-125723.9 ABb
are also reproducing well the spectrophotometry of HD~95086\,b.

We compare in Fig.~\ref{Fig:visufit} the spectrophotometry of
HD~95086\,b to some of the best fitting templates. The planet has an
$H$-band flux (SPHERE and GPI) lower than the one of the isolated
objects. This discrepancy is slightly reduced for the youngest
objects. The companion VHS J125601.92-125723.9 ABb appears to produce
the best fit to the 1.2-2.3 $\mu$m spectrophotometry of the
HD~95086\,b. While the peculiar L9 dwarf WISE1647+56 occupies the same
location as HD~95086\,b in the color-magnitude diagrams
(Fig.~\ref{Fig:CMD}), this object has a bluer spectral slope than the
companion.

We compared in Fig.~\ref{Fig:SED} the 1-4 $\mu$m spectrophotometry
to those of objects with published $L\,'$ photometry or $L$-band spectra
\citep{2004A&A...425L..29C, marois2008, 2009ApJ...702..154S,
  2016ApJ...830..114R} and known distances \citep{2008A&A...477L...1D,
  2012ApJ...752...56F, 2016A&A...595A...2G, 2016ApJ...833...96L,
  2015ApJ...804...96G,
  2016ApJ...829...39S}.\footnote{\cite{derosa2016} performed an
  analysis which includes many more comparison objects without
  published $L\,'$ band photometry. To do so, they converted the WISE 1
  magnitude to a $L\,'$ band magnitude using empirical relations based on
  the use of empirical spectra from the IRTF Spectral Library and the
  SpeX Prism Spectral Library. We prefered to avoid using those
  conversion formulae because the empirical libraries considered to
  derive them contain mostly spectra of old objects. Yet, the L-band
  shape is known to be influenced by the age/surface gravity
  \citep[see][]{2009ApJ...702..154S, 2016ApJ...829...39S}.} These reference spectra 
were re-normalized following Eq. \ref{eq:gk} and \ref{eq:alpha} to minimize $G$ when compared to the 1-4 $\mu$m spectral energy distribution of HD~95086\,b. This comparison confirms that the planet $H$
band flux is fainter compared to the average flux of the templates. In
addition, the spectral slope of HD~95086\,b is much redder than those
objects. This confirms the conclusions of \cite{derosa2016}.  The
slope of the 1-4 $\mu$m SED of the L7 dwarf 2MASS J22443167+2043433
(AB Dor candidate member) is redder than the one of the L7.5 Hyades
candidate member 2MASS J08251968+2115521
\citep{2007MNRAS.378L..24B}. Since HD~95086\,b is much younger than AB
Dor, this suggests that its very red slope is the extreme example of
the impact of surface gravity on the photospheric opacities.
\cite{2016ApJ...833...96L} showed that the underluminosity of young
L/T transition objects compared to field objects is more dramatic in
the $Y$, $J$, and $H$ bands while it is reduced in the $K$ and $W1$
bands. Because HD~95086\,b is an extreme object, we believe that the
bolometric corrections, defined by \cite{2015ApJ...810..158F} for the
$J$ and $K$ bands and for young objects, are not suitable for the planet and
should not be used to derive the luminosity.

In summary, the empirical analysis reveals that the planet 1-2.5
$\mu$m photometry is represented by a few empirical objects with
estimated spectral types L7-L9, which are candidate members of young
moving groups, and whose masses are close to the one estimated for the
planet ($4-5$\,M$\mathrm{_{Jup}}$). The extremely red 1-4 $\mu$m spectral energy distribution of
HD~95086\,b and its underluminosity are characteristics of low gravity
objects of the L/T transition and render the estimate of the bolometric luminosity uncertain.  
%The planet HD~131399ABb \citep{2016Sci...353..673W}
%is estimated to have a mass and an age in the same range as
%HD~95086\,b. But this object does not show an obvious underluminosity in
%the color-magnitude diagrams (Fig. \ref{Fig:CMD}) and has a methane
%absorption typical of T dwarfs. This shows that the L/T transition for
%objects in the 10-20 Myr range is sharp. This also means that
HD~95086\,b probably falls in a regime where the underluminosity related
to young objects is maximum, especially at $JHK$ bands where GPI and
SPHERE operate. In comparison, HD~95086\,b has the same $L$-band
luminosity as 2M1207b and HR8799c and d \citep{galicher2014}, but has
fainter fluxes at shorter wavelengths \citep[this work
  and][]{derosa2016}. Therefore, complementary $L\,'$ band observations of
Sco-Cen targets should be obtained to ensure that planets falling
right at the L/T transition are not missed.

%%%%%%%%%%%%%%%%%%%%%%%%%%%%%%%%%%%%%%%%%%%%%%%%%%%%%%%%%%%%%
\section{Constraints on additional giant planets}

The observations of HD\,95086 with HARPS and SPHERE did not reveal the
presence of new planets in the system in addition to
HD\,95086\,b, which is well resolved with both IRDIS and IFS (as
described in Section~2, see Fig.~\,\ref{fig:image} and
\ref{fig:ifs}). The combination of all detection limits can however enable us
to constrain the completeness of our study to identify the parameter
space in mass and semi-major axis for which planets are
detected. After describing below the approaches used to derive the
individual detection limits for HARPS and SPHERE, we detail the
outcome of a set of Monte-Carlo simulations to probe the probable
presence of any additional planets in this young planetary system from
a few stellar radii up to 800~au.

\subsection{Combined HARPS and SPHERE detection limits}

For HARPS measurements, we used the local power analysis (LPA)
developed by \cite{meunier2012}. This method is based on the
generation of periodograms of synthetic planet RV time series, that
are compared with the periodogram of the observed RV data within given
orbital periods. A detection requires the planet-induced power of the 
RV signal to be higher than the power of the actual signal within a localized 
period range. The exploration in mass and period (or semi-major axis) enable to set the detection 
limit in terms of detection probablity. Please refer to Meunier et al. (2012) 
for a detailed description of the LPA method in comparison with the standard rms or
correlation-based approaches \citep{meunier2012}. 

For IRDIS, a standard pixel-to-pixel noise map of each observation was
estimated within a box of 5$\times$5 pixels sliding from the star to
the limit of the IRDIS and IFS field of view. To correct for the flux
loss related to the ADI processing, fake planets were regularly
injected every 20~pixels in radius at 10 different position angles for
separations smaller than $3~\!''$. At more than $3~\!''$, fake planets
were injected every 50~pixels at 4 different position angles. The
final flux loss was computed with the azimuthal average of the flux
losses of fake planets at same radii. For IFS, noise maps were also
estimated taking account the flux loss with the injection of fake
planets, (flat spectra), then summed over the red-part between 1.5 and
1.66~$\mu$m. The final detection limits at $5\sigma$ were then
obtained using the pixel-to-pixel noise map divided by the flux loss
and normalized by the relative calibration with the primary star
(considering the different exposure times, the neutral
density and the coronograph transmission). These detection limits in contrast were finally corrected from small number statistics following the prescription of Mawet et al. (2014) to adapt our $5\sigma$ confidence
level at small angles with IRDIS and IFS. The results for IRDIS and
IFS are reported in Fig.~\ref{fig:detlim}.

\begin{figure*}[th]
\hspace{0.2cm}
\includegraphics[width=\columnwidth]{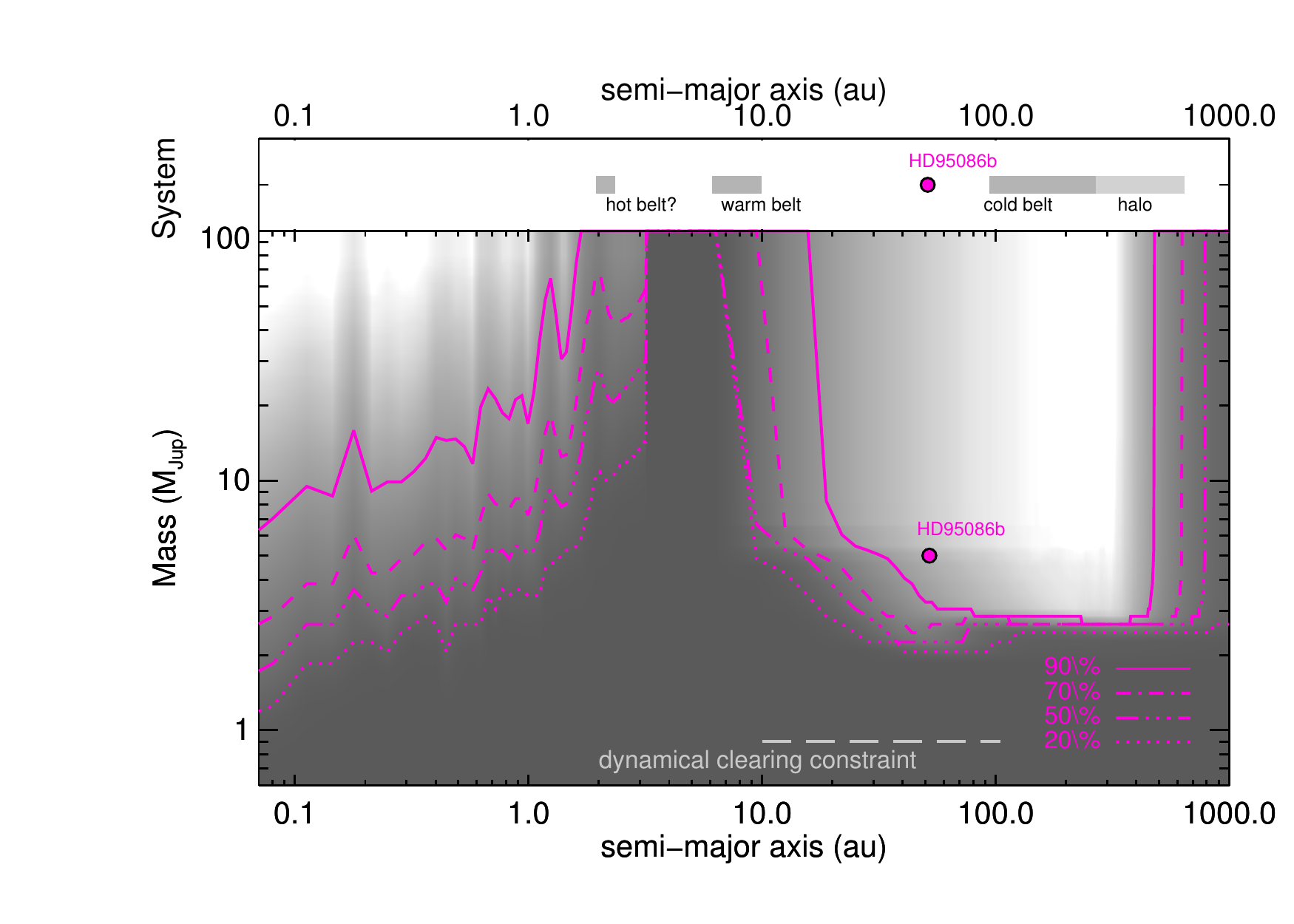}
\includegraphics[width=\columnwidth]{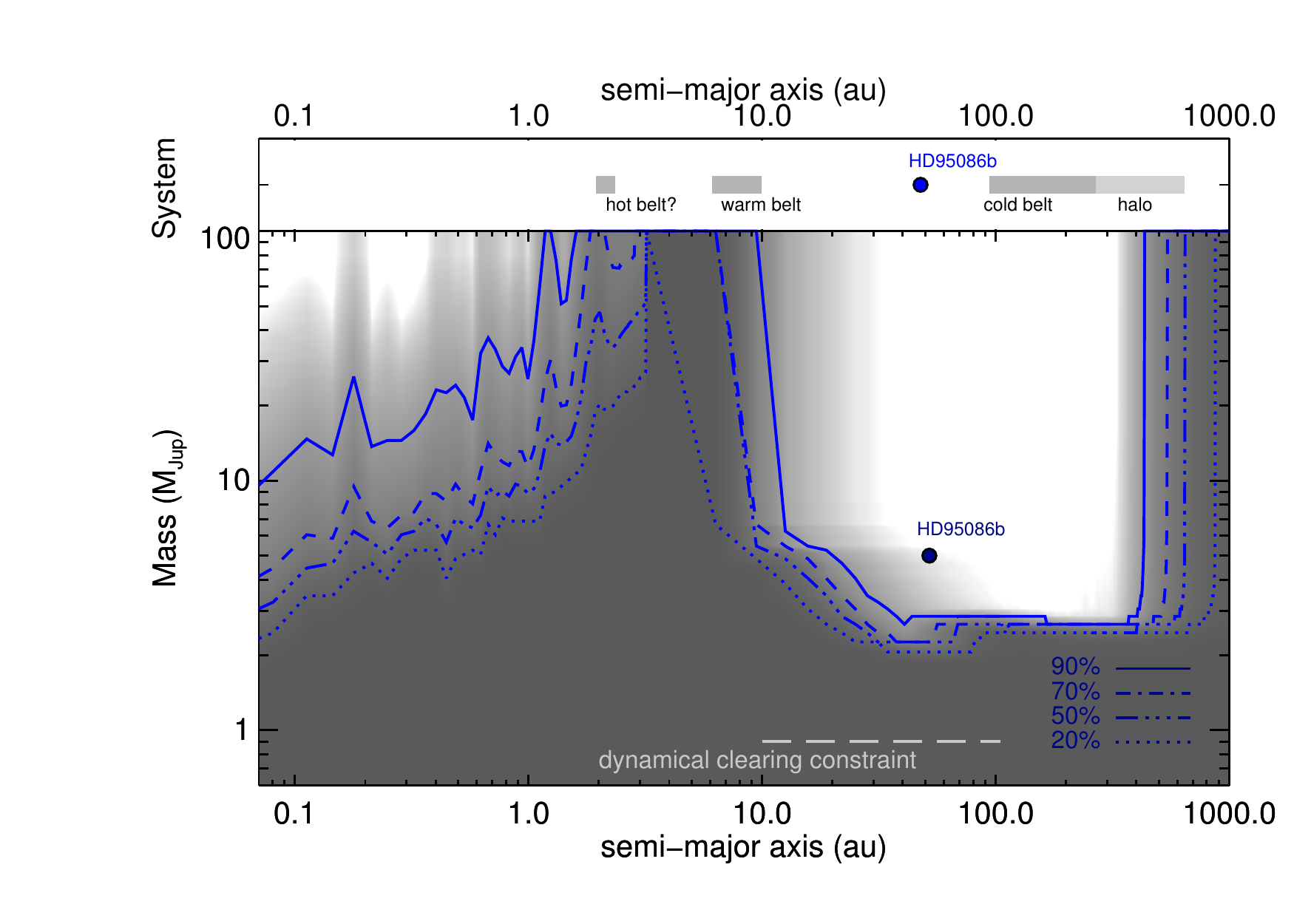}
\begin{centering}

\caption{HARPS and SPHERE combined detection probabilities given as a
  function of the planet mass and semi-major axis. \textit{Left:}
  Detection probabilities given for planet orbits with no constraints
  on the inclination, i.e. randomly oriented in space with uniform
  distributions in $cos(i)$. Together with HD\,95086\,b, the dust
  distribution in the system coming from the combined analysis of the
  SED fitting and resolved PACS far-IR images is reported with the
  location of the three components: \textit{warm}, \textit{cold},
  \textit{halo} in addition to the possible \textit{hot} one suggested
  at 2~au.  \textit{Right:} Detection probabilities with constraints
  set to restrain solutions to planet orbital inclinations coplanar
  with the disk inclination of $i=-30\pm3\degr$ and $i=150\pm3\degr$. }

\label{fig:mess2}
\end{centering}
\end{figure*}

\subsection{Planet detection probability around HD\,95086}

To combine the HARPS and SPHERE detection limits, we used an optimized
version of the MESS (multi-purpose exoplanet simulation system) code,
a Monte Carlo tool for the statistical analysis and prediction of
exoplanet search results \citep{bonavita2012}. The latest MESS2
version can now handle the combination of radial velocity and direct
imaging measurements obtained at various epochs. Please refer to \cite{lannier2017} 
for a detailed description of the code evolution. 

In the case of HD\,95086, we used MESS2 to 
generate a uniform grid of mass and semi-major axis in the interval [1, 80]~M$_{\rm{Jup}}$ and [1, 1000]~au with a
sampling of 0.5~M$_{\rm{Jup}}$ and 1~au, respectively. The same synthetic planet population 
is used to explore the detection probability of both HARPS and SPHERE measurements.
For each point in the grids, 100 orbits were generated, randomly oriented in space
from uniform distributions in cos(i), $\omega$, $\Omega$, $e \le 0.8$,
and $T_p$. Following the LPA approach described before, the Lomb-Scargle periodograms \citep{lomb1976,scargle1982} 
are computed for each generated planet, as well as the periodogram of the observed RV data. Detection probability map are built by counting the number of detected planets over the number of generated ones. The detection threshold is defined when the maximum planet-induced power of the periodogramme is 1.3 times larger than the maximum power of the observed RV data within the period range \citep{meunier2012}. In addition, the on-sky projected position (separation and position
angle) at the time of the observation is computed for each planetary orbit
and compared to the SPHERE \textbf{$5\sigma$} 2D-contrast detection maps 
converted in masses based on the COND model predictions \citep{baraffe2003}. The primary age, distance and magnitude, reported in Table\,\ref{prop}, are considered for the luminosity-mass conversion. Detection probablities are built from 
the number of detected planets over the number of generated ones given our direct-imaging detection
threshold.

The MESS2 results merging both HARPS and SPHERE detection probabilities are shown
in Fig.~\ref{fig:mess2} \textit{left-panel} when no specific orbital
constraints are set on the planet properties. Although there is no clear overlap between the radial velocity and
direct imaging techniques, the combination of both allows the
exploration of giant planets with semi-major axis smaller than 3~au
and masses typically larger than 7~M$_\mathrm{Jup}$ in the best cases
and for semi-major axes between typically 15 to 500~au and masses
larger than 2~M$_\mathrm{Jup}$, respectively for detection probability
of 90\%. For detection probability of 20\%, these values decrease to
masses larger than 1~M$_\mathrm{Jup}$ in the best cases for radial velocity
at less than 3~au and 2~M$_\mathrm{Jup}$ between 10 and 1000~au in
direct imaging. On the \textit{Right-Panel} of Fig.~\ref{fig:mess2}
we report the detection probabilities considering planetary orbital
solutions that are coplanar with the disk inclination ($i=-30\pm3$~$\degr$
and $i=150\pm3$~$\degr$). These close to face-on
solutions logically favor deeper exploration of the close-in physical
separations in direct imaging whereas detection probabilities at short
periods with radial velocity become less sensitive. In the most
optimistic cases, detection probabilities go down to
2~M$_\mathrm{Jup}$ in radial velocity at less than 3~au and
2~M$_\mathrm{Jup}$ between 10 and 1000~au in direct imaging for
detection probability of 20\%.  Forthcoming \textit{Gaia} astrometric results on this system will complement 
the current radial velocity and direct imaging constraints on the population of 
massive jovian planets by accessing a typical 
discovery window between 1 to 10 au.

%%%%%%%%%%%%%%%%%%%%%%%%%%%%%%%%%%%%%%%%%%%%%%%%%%%%%%%%%%%%%%%%%%%%%%
\section{Polarized-light detection of the outer disk}
\label{sect:disk}

The main properties of the dust distribution studied by \cite{moor2013}, \cite{su2015} and \cite{su2017} combining the analysis of the disk SED, PACS far-infrared images and recent ALMA 1.3\,mm observations are summarized in
Table~\ref{di}. The detail of the dust temperature and predicted location is given for the four \textit{hot},
\textit{warm}, \textit{cold} and \textit{halo} components around
HD\,95086. A key result of the ALMA map is to resolve for the first time the 1.3\,mm emission of the \textit{cold} outer belt, which is consistent with a broad ring with sharp boundaries from $106\pm6$\,au to $320\pm20$\,au.

\begin{figure}[t]
\hspace{0.2cm}
\includegraphics[width=\columnwidth]{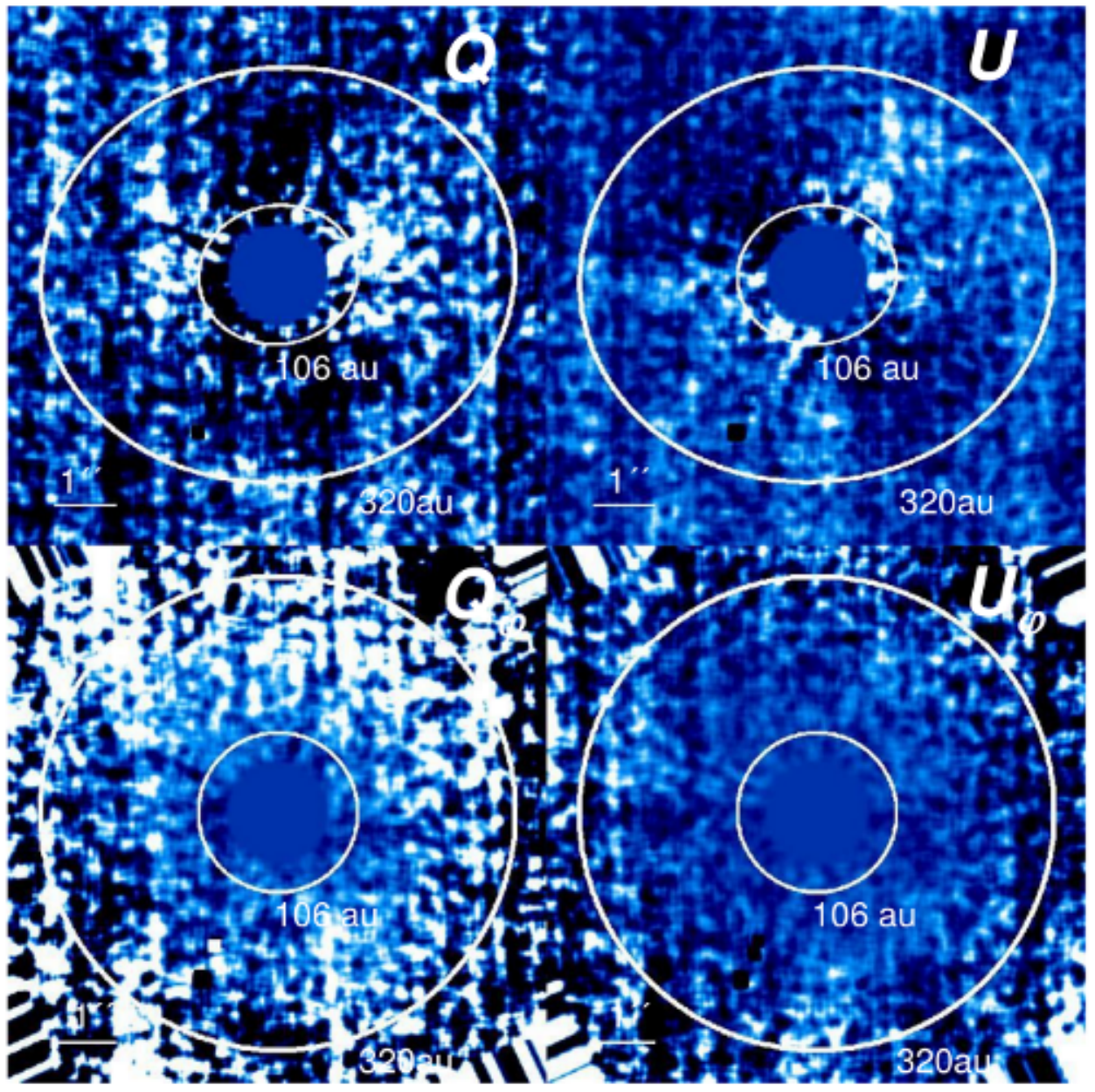}
\begin{centering}

\caption{\textit{Upper-Left:} IRDIS $J$-band image of the Q Stokes parameter.
\textit{Upper-Right:} Same for the U Stokes parameter. North is up and East is toward the Left. The location of the inner (106\,au) and outer (320\,au) edges 
of the \textit{cold} outer component is superimposed in each case. \textit{Bottom-Left:} IRDIS $J$-band image of the Q$_\phi$ radial polarized Stokes parameter. \textit{Bottom-Right:} Same for the U$_\phi$ radial polarized Stokes parameter. Both images were corrected for the 
disk inclination and position angle and scaled 
with $r^2$ to compensate for the $r^{-2}$ dependency on the stellar 
illumination. Deprojected location of the inner and outer radii 
of the \textit{cold} outer component is indicated.}

\label{fig:disk1}
\end{centering}
\end{figure}

A prime objective of the IRDIS-DPI observations was to image the disk outer part in
$J$-band polarized scattered light at sub-arcsecond resolution, taking advantage of the
spatial resolution achievable with IRDIS (145~mas-diameter for the
coronagraphic mask i.e. down to 10~au). The outcome of the DPI data
processing of HD\,95086, the Q and U Stokes parameters and the Q$_\phi$ and U$_\phi$ radial polarized
Stokes parameters (see description in sub-Sect.~2.1.1), 
is reported in Fig.~\ref{fig:disk1}. Both Q$_\phi$ and U$_\phi$ images were corrected for the 
disk inclination and position angle, assuming an a-priori knowledge on the disk geometry, and scaled 
with $r^2$ to compensate for the $r^{-2}$ dependency on the stellar 
illumination. No corrections were applied to both Q and U images. Although no obvious signal is detected in the Q$_\phi$ image compared to U$_\phi$, there is a clear Q and U butterfly detection at the location of the \textit{cold} belt and \textit{halo} components fitted by 
\cite{su2017} (from 106 to 320~au, i.e. $1.2\,''$ to $3.8\,''$). This can be explained by the fact that the oscillating pattern of the Q and U butterfly (which behaves as expected) clearly stands out against the noise, while for an all positive Q$_\phi$ or polarized intensity image it is more difficult.

\begin{table}[t]

\caption[]{Disk properties for the \textit{hot}, \textit{warm}, \textit{cold} and \textit{halo} components
  identified in that system based on \cite{su2015} and updated values of \cite{su2017}. Parameters refer to the temperature of dust
  ($T_{\rm{dust}}$), radial location, inclination, and position angle (P.A.).}
\label{di}
\centering                          % used for centering table
\begin{tabular}{cccccc}
\noalign{\smallskip}\hline\noalign{\smallskip}
Parameter & Unit &  \textit{Hot?} & \textit{Warm} & \textit{Cold} & \textit{Halo} \\

\noalign{\smallskip}\hline\noalign{\smallskip}

$T_{\rm{dust}}$  & [K]         & $300$     & 175     & 55       & - \\
\noalign{\smallskip}
$r_{\rm{dust}}$  & [AU]        & $\sim2$ & $7-10 $ & $106-320$ & $300-800$ \\
\noalign{\smallskip}
%$f_{\rm{dust}}$ ($10^{-3}$) &   &   -     & 0.15    & 1.5      & $\sim0.3$ \\
%\noalign{\smallskip}
%$M_{\rm{dust}}$  & [$M_{\oplus}$]&   -    & 4-5e-5     & 0.2   & 0.02-0.05\\
\noalign{\smallskip}
Inclination   &  [$^o$]           &   -    &       -  &           $30\pm3$   & -      \\
\noalign{\smallskip}
P.A. & [$^o$]           &   -    &    -     &           $97\pm3$  & -       \\
%\noalign{\smallskip}
%Eccentricity&                   &  -     &    -     &           $<0.3$  & -\\
\noalign{\smallskip}\hline
\end{tabular}
\end{table}

\begin{figure}[t]
\hspace{0.2cm}
\includegraphics[width=\columnwidth]{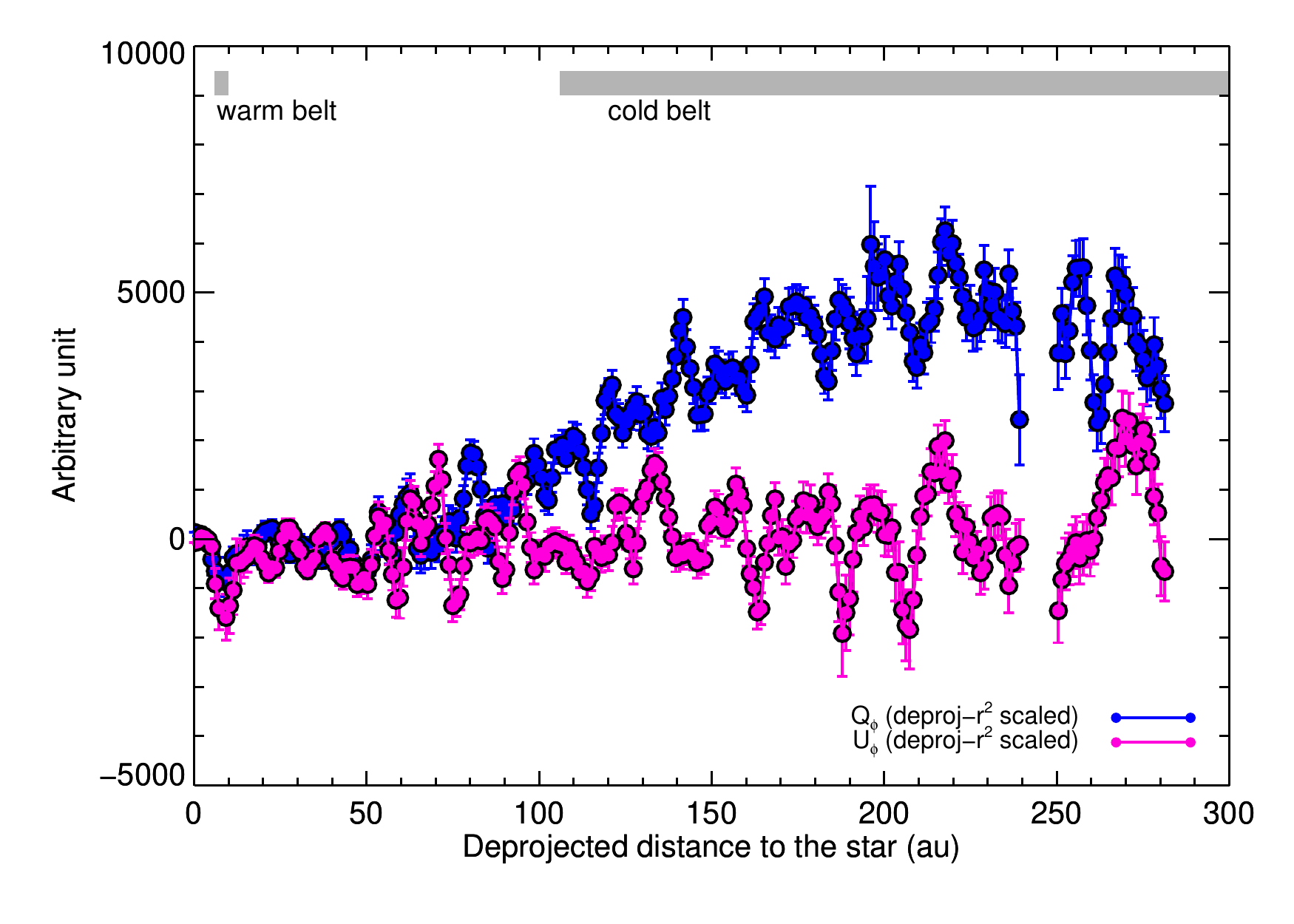}
\begin{centering}

\caption{Azimuthal average of the Q$_\phi$, U$_\phi$ radial polarized
  Stokes parameters calculated with error bars. Both parameters were corrected for the 
disk inclination and position angle and scaled 
with $r^2$ to compensate for the $r^{-2}$ dependency on the stellar 
illumination. The location of the 
\textit{warm} inner, \textit{cold} outer and \textit{halo} components are indicated.}

\label{fig:disk2}
\end{centering}
\end{figure}

%\begin{table}[t]

%\caption[]{Disk properties for the \textit{hot}, \textit{warm}, \textit{cold} and \textit{halo} components
%  identified in that system based on \cite{su2015}. Parameters refer to the temperature of dust
%  ($T_{\rm{dust}}$), radial location, fraction of total dust
%  luminosity ($f_{\rm{dust}}$) and dust mass ($M_{\rm{dust}}$).}
%
%\label{di}
%\centering                          % used for centering table
%\begin{tabular}{cccccc}
%Parameter & Unit &  \textit{Hot?} & \textit{Warm} & \textit{Cold} & \textit{Halo} \\%


%$T_{\rm{dust}}$  & [K]         & $300$     & 175     & 55       & - \\
%\noalign{\smallskip}
%$r_{\rm{dust}}$  & [AU]        & $\sim2$ & $7-10 $ & $63-190$ & $\lesssim800$ \\
%\noalign{\smallskip}
%$f_{\rm{dust}}$ ($10^{-3}$) &   &   -     & 0.15    & 1.5      & $\sim0.3$ \\
%\noalign{\smallskip}
%$M_{\rm{dust}}$  & [$M_{\oplus}$]&   -    & 4-5e-5     & 0.2   & 0.02-0.05\\
%\noalign{\smallskip}
%Inclination   &  [$^o$]           &   -    &       -  &           $25\pm5$   & -      \\
%\noalign{\smallskip}
%P.A. & [$^o$]           &   -    &    -     &           $115\pm10$  & -       \\
%\noalign{\smallskip}
%Eccentricity&                   &  -     &    -     &           $<0.3$  & -\\
%\end{tabular}
%\end{table}

%\subsection{Planet-disk connection}

% Warm component origin?
% Need of additional planets in the system?

To further explore the presence of a spatially resolved and faint signal in the Q$_\phi$ image, 
we calculated the azimuthal average as a function of the radial extent together with the
related uncertainties quadratically normalized by the number of
elements in each azimuthal ring. The result is reported in
Fig.~\ref{fig:disk2}. A faint diluted signal emerges from a projected radius of $1.2\,''$ up to $3.5\,''$ (cutoff of sensitivity in our IRDIS DPI observations). It
corresponds to deprojected physical separations of 100 and 300~au,
respectively. Both pipelines independently confirm the observed over-intensity in Q$_\phi$
compared to U$_\phi$. Although faint, this detection enables for the first time to resolve the polarized scattered light from the \textit{cold} outer
component (and possibly the extended \textit{halo}) surrounding HD\,95086. The detection shows a drop of polarized flux below $1.2\,''$ ($\sim100$\,au)
compatible with the presence of a large cavity between typically 10 and
100~au resulting from the multi-belts analysis combining disk SED,
PACS and ALMA images. It also fits with the presence of the closer-in HD\,95086\,b exoplanet at a semi-major axis of roughly 52\,au. Finally, our result corroborates the radial extent of the \textit{cold} outer belt lying between 100\,au up to at least 300\,au as observed with ALMA.

%\begin{figure}[t]

%%%%%%%%%%%%%%%%%%%%%%%%%%%%%%%%%%%%%%%%%%%%%%%%%%%%%%%%%%%%%
\section{Planets and belts}

The young planetary system around HD\,95086 offers an interesting
comparison case with HR\,8799 \citep{gotberg2016,booth2016,konopacky2016,su2015}, and more
generally with the interpretation that these multiple-belt debris
disks are actually young analogues to our Solar system. The presence
of planets would be responsible for the dynamical clearing of the
debris disk and the formation of observed multi-belt architecture as
suggested by \cite{kennedy2014} and \cite{shannon2016}. In the specific case of HD\,95086,
one could actually wonder if the presence of additional planets to
HD\,95086\,b would be necessary to explain the observed
planet - belt architecture including: i/ a \textit{warm} and
relatively narrow inner belt at $\sim8\,$au, ii/ a broad cavity from
typically 10 to 100\,au inside which orbits the massive (4--5~$M_{\rm{Jup}}$, $a\sim52_{-24.3}^{+12.8}$\,au and $e=0.2_{-0.2}^{+0.3}$) planet HD\,95086\,b, iii/ finally, a \textit{cold}
outer belt lying between 106 to 320\,au extending within a disk \textit{halo}
component up to 800\,au.

\begin{figure}[t]
\hspace{0.2cm}
\includegraphics[width=\columnwidth]{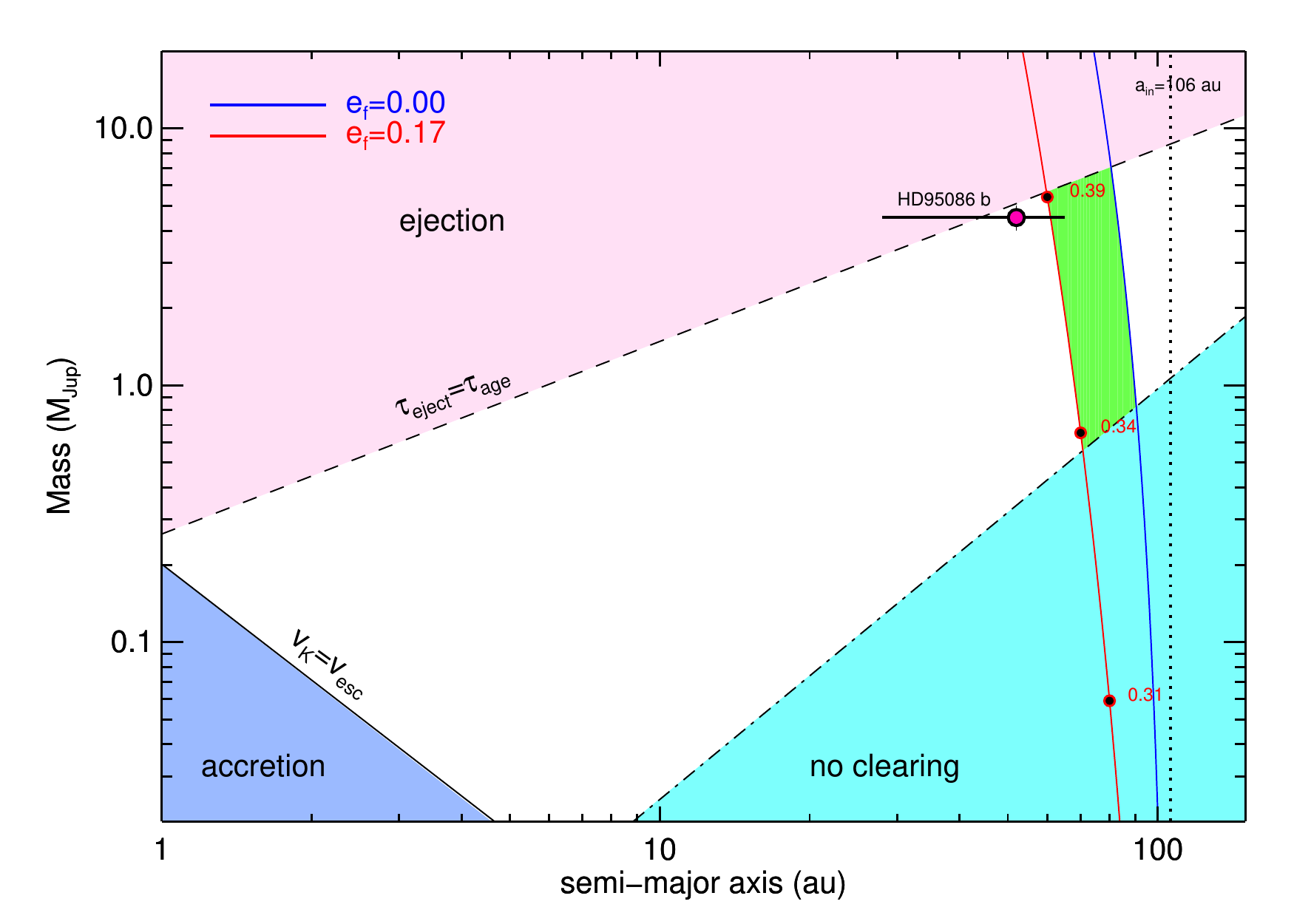}
\begin{centering}
\caption{Constraints on the mass, semi-major axis and eccentricity of the planetary perturber to explain  both the inner belt replenishment and the formation and eccentricity upper limit of the outer belt. \textit{Light purple} and \textit{dark blue} areas show excluded regions where particles encountering a planetary perturber of given semi-major axis and mass will not remain in the system to resplenish the inner belt without being ejected or accreted for timescales younger than the age of the system, respectively. Planets with escape velocity, $V_{esc}$, larger than the Keplerian velocity,
$V_K$, will most likely eject particles after multiple kicks. On
the other hand, if $V_{esc} \ll V_K$ then accretion will likely be the
final outcome. This sets the limit shown by the \textit{solid black line}. In addition, as ejection might only happen after several encounters, thus,
material can remain in the system for timescales shorter
than the ejection timescale. Considering HD\,95086's age, this set the ejection limit shown with the (\textit{dashed black line}). Moreover, \textit{light blue} area shows region where the planetary perturber will be not massive enough to stir the disk and carve the outer belt. The \textit{blue} and \textit{red} curves delimit dynamical solutions for which the planetary perturber can create an outer belt eccentricity ranging between 0.00 (circular belt) and 0.17 (ALMA upper limit from \citealt{su2017}), respectively. Corresponding planetary perturber eccentricities required to create an outer belt eccentricity of 0.17 are overplotted in \textit{red}. The \textit{green} area shows the final region for with the planetary perturber semi-major axis, mass and eccentricity meet all above dynamical constraints to explain both the the inner belt replenishment and the formation and eccentricity upper limit of the outer belt. Finally, the predicted location of HD\,95086\,b by our MCMC orbital fitting analysis is reported in \textit{pink} with error bars.}
\label{fig:mar}
\end{centering}
\end{figure}

Given that dust has a short lifetime against collisions and 
radiative forces, the population of small grains in the \textit{warm} inner belt observed in the spectral energy distribution of HD\,95086 must be continuously replenished.
\cite{su2015} rejected P-R drag as a dominant mechanism to
transport dust from the outer belt toward the star. They found a
maximum amount of material dragged one
order of magnitude below what is observed in fractional
luminosity for the \textit{warm} inner belt. An alternative mechanism 
proposed by \cite{bonsor2012} to sustain exozodiacal dust originates from small bodies
of an outer belt, scattered inwards by planets. Their N-body simulations show that a 
chain of closely spaced, Saturn to Jupiter-mass planets orbiting interior to a massive outer 
planetesimal belt could replenish exozodiacal dust and sustain an asteroid belt analog. 

\cite{marino2017} applied this scenario to 
the emblematic case of the 1-2 Gyr old star $\eta$ Corvi (F2V, 18.3\,pc) surrounded by a hot dust component located at $\sim1.4$\,au and a colder one at $\sim150$\,au.
Considerations on the planet and disk eccentricities,
clearing timescale of the chaotic zone as a function of the planet
mass and semi-major axis, and the migrating particle escape velocity
when encountering the perturbing planet led them to
constrain the most probable mass, semi-major axis and eccentricity for the
perturbing planet. We applied the same formalism (Eq. 15, 16, 17, 18 and Eq. 19 of \citealt{marino2017}) 
to the younger case of HD\,95086 (1.6\,$M_{\star}$, $17\pm2$\,Myr) surrounded by an outer belt located at 200\,au with an inner boundary at 106\,au. We considered in addition an upper limit for the outer belt eccentricity of $e<0.17$
coming from the non-detection of an outer belt offset in the ALMA 1.3\,mm 
resolved observations of \cite{su2017}. The results are shown in Fig.\,\ref{fig:mar}. 

The \textit{light purple} and \textit{dark blue} areas show excluded regions where particles encountering a planetary perturber will not remain in the system to resplenish the inner belt without being ejected or accreted for timescales younger than the age of the system, respectively. The \textit{light blue} area shows region where the planetary perturber would not be massive enough to 
stir the disk and carve the outer belt. The resulting \textit{white} area therefore defines the parameter space for which the planetary perturber will be able to carve the outer belt and resplenish the inner belt.

If we add in addition constraints from the outer belt eccentricity obtained by ALMA, the \textit{blue} and \textit{red} curves delimit dynamical solutions for which the planetary perturber can create an outer belt eccentricity ranging between 0.00 (circular belt) and 0.17 (ALMA upper limit from \citealt{su2017}), respectively. The discrete planetary perturber eccentricities required to create an outer belt eccentricity of 0.17 are indicated in \textit{red}. The \textit{green} area shows the final region for the planetary perturber semi-major axis, mass and excentricity where all above dynamical constraints are met to explain both the the inner belt replenishment and the formation and eccentricity upper limit of the outer belt.

If we report the orbital properties of HD\,95086\,b ($a=52_{-24.3}^{+12.8}$\,au and $e=0.2_{-0.2}^{+0.3}$) and its predicted mass (4--5~$M_{\rm{Jup}}$), we see that HD\,95086\,b falls at a location consistent with the predicted properties of the dynamical perturber sculpting the inner boundary of the \textit{cold} outer component, but for moderate-eccentricity ($e=0.3-0.4$) solutions. If further astrometric monitoring rather indicate low-eccentricity solutions for HD\,95096\,b, an additional outer planet at larger semi-major and with masses below $\sim2$~$M_{\rm{Jup}}$ (given the current detection limits) will have to be considered to explain the observed architecture as proposed by \cite{su2017}. 

The broad cavity observed between 10 and 100\,au and the current orbital properties of HD\,95086\,b constrained by GPI and SPHERE suggest anyway that the presence of additional planets is necessary to clear the cavity. In the case of multiple-belt debris disk, \cite{shannon2016} explored with N-body simulations the minimum planet mass and the expected number 
of planets that must be present to produce a broad cavity for a star of a given age. If we apply Eq. 4 and 5 of \cite{shannon2016} to the case of HD\,95086 (1.6\,$M_{\star}$, $17\pm2$\,Myr, cavity from 10 to 100\,au), we derive a minimum mass of the planets in the cavity of $0.35\,M_{\rm{Jup}}$ and a typical number of requested planets of 2.4 (i.e. 2 to 3 giant planets depending on their respective separation). If we compare these results to the present outcome of the HARPS and SPHERE combined detection limits (Fig.~\ref{fig:mess2}, \textit{Right}, coplanar case), we see that there might still be room for 2 additional stable planets c and d in the cavity in addition to b with typical masses between $0.35\,M_{\rm{Jup}}$ (dynamical clearing constraint) and $6\,M_{\rm{Jup}}$ for semi-major axis between 10 and 30\,au or $0.35\,M_{\rm{Jup}}$ and $5\,M_{\rm{Jup}}$ beyond 30\,au.  If we rule out the presence of additional planets at $\sim8$~au owing to the presence of the \textit{warm} inner belt, closer-in planets might of course exist but with less constraining limits in mass of typically $10$ to $20\,M_{\rm{Jup}}$ if orbiting inside 3\,au according to our HARPS sensitivity curves. Further N-body
simulations tuned for this system would help to refine these conclusions. Although the
combination of HARPS and SPHERE in this specific case clearly
illustrates the gain of combining techniques (radial velocity and direct
imaging) to probe the presence of close-in and wide orbits giant
planets in single systems, we see that the new generation of extremely
large telescopes in combination with radial velocity and astrometric surveys
will be necessary for a global and full exploration of the giant
planet population around systems at distances larger than typically 100 pc.

\section{Conclusions}

In the course of the HARPS large program targeting young, nearby
stars and the SHINE and DISK SPHERE GTO programs, we observed the
young, planetary system around HD\,95086 to explore its global
architecture. Our prime goals were to image the \textit{cold} or \textit{halo} outer
component of the HD\,95086 debris disk resolved by \textit{Herschel}
in far-IR and recently ALMA at 1.3\,mm, to characterize the physical properties of the known imaged
planet HD\,95086\,b (orbital and atmospheric properties), finally to
search for additional planets, constrain their possible physical
properties and discuss them in the light of the formation and viable
dynamical configurations of the multi-belt architecture observed in
that system. Here, we summarize the main results:

\begin{enumerate}

\item we do not detect any additional planet in the system with
  either HARPS or SPHERE. The 10 point-like sources detected in the
  SPHERE/IRDIS field of view of $12.5\,''\times12.5\,''$ in addition
  to HD\,95086\,b are all identified as background objects.

\item HD\,95086\,b is well resolved with IRDIS in $K1$ and $K2-$bands
  at four epochs between February 2015 and May 2017. The planet's
  orbital motion is unambiguously resolved. The results of our MCMC
  orbital fitting analysis favor retrogade orbital solutions of
  low- to moderate-eccentricity $e\la0.5$, with a semi-major axis $\sim52$~au
  corresponding to orbital periods of $\sim288$\,yrs and an
  inclination that peaks at $i=140.7\,\degr$, which are still
  compatible with a planet-disk coplanar configuration.

\item HD\,95086\,b is imaged at $H$-band and for the first time at
  $J$-band using the SPHERE IFS instrument by stacking reduced images
  taken at various epochs to optimize the speckle cancellation. Its
  near-infrared spectral energy distribution is well fitted by a few
  dusty and/or young L7-L9 dwarfs. It shows an extremely red $1-4$\,$\mu$m
  spectral distribution typical of low-gravity effect in the atmospheres of young
  exoplanets at the L/T transition.

\item the combination of HARPS and SPHERE detection limits
  offer the unique possibility to explore and constrain the
  physical properties of additional giant planets in HD\,95086 at
  close-in and wide orbits. Although there is no clear overlap between
  both techniques, we reject in the most optimistic cases (detection
  probability of 20\%) and for a planet-disk coplanar configuration
  the presence of giant planets with masses larger than 2~M$_\mathrm{Jup}$
  to 10~M$_\mathrm{Jup}$ at less than 3~au and masses larger than 5~M$_\mathrm{Jup}$ between 10 and 30\,au 
  and larger than 2~M$_\mathrm{Jup}$ beyond 30~au. 

\item finally, the outer debris belt around HD\,95086 is resolved for the first time in polarized scattered light by
  our IRDIS $J$-band differential polarimetric imaging. The
  radial extent of the detected diffused polarized flux is compatible
  with the location of the \textit{cold} component recently resolved by ALMA 1.3\,mm observations.

\item These results enable us to discuss the presence of additional planets to HD\,95086\,b in that 
  system to explain the replenishment of \textit{warm} inner belt located at $\sim8~$au, the origin of  
  the broad cavity extending between 10 to 100\,au, and the observed properties of the \textit{cold} outer belt beyond 100\,au. 
  They illustrate the rich synergy offered
  by the combination of various observing techniques to explore
  the global content of giant planets aroung young, nearby stars and their ability to shape planetary system architectures.
 
\end{enumerate} 

The young planetary system  HD\,95086 has undoubtedly become one of this rare, enblematic laboratory for the study of giant planet formation as HR\,8799 \citep{marois2010,2016A&A...587A..57Z,booth2016}, $\beta$ Pictoris \citep{lagrange2010,dent2014,wang2016}, HR\,4796 \citep{perrin2015,milli2017}, HD\,61005 \citep{olofsson2016,esposito2016}, Fomalhaut \citep{kalas2008,macgregor2017}, AU\,Mic \citep{boccaletti2015,wang2015}, or TW\,Hya \citep{rapson2015,vanboekel2017} and HL\,Tau \citep{alma2015} at younger ages. HD\,95086 will remain a prime target for the SPHERE and GPI planet imagers and high-resolution spectrographs like HARPS in the coming decade to further explore the presence of additional planets in the system, soon for \textit{JWST} from space, and for the versatile instrumentation of extremely large telescopes that will bridge the gaps between the different observing techniques to further explore the diversity of this young Solar System analog.

%\subsection{HD\,95086\,b orbit determination}
%\subsection{HARPS + SPHERE combination}
%\subsection{Disk modeling + planet}
%\subsection{GPI/SPHERE comparison}
%------------------------------------------------------------------------
\bibliographystyle{aa}

\begin{acknowledgements}

We acknowledge financial support from the Programme National de
Planétologie (PNP) and the Programme National de Physique Stellaire
(PNPS) of CNRS-INSU. This work has also been supported by a grant from
the French Labex OSUG@2020 (Investissements d’avenir – ANR10 LABX56).
The project is supported by CNRS, by the Agence Nationale de la
Recherche (ANR-14-CE33-0018). This work is partly based on data
products produced at the SPHERE Data Centre hosted at OSUG/IPAG,
Grenoble. We thank P. Delorme and E. Lagadec (SPHERE Data Centre) for
their efficient help during the data reduction process. SPHERE is an
instrument designed and built by a consortium consisting of IPAG
(Grenoble, France), MPIA (Heidelberg, Germany), LAM (Marseille,
France), LESIA (Paris, France), Laboratoire Lagrange (Nice, France),
INAF–Osservatorio di Padova (Italy), Observatoire de Genève
(Switzerland), ETH Zurich (Switzerland), NOVA (Netherlands), ONERA
(France) and ASTRON (Netherlands) in collaboration with ESO. SPHERE
was funded by ESO, with additional contributions from CNRS (France),
MPIA (Germany), INAF (Italy), FINES (Switzerland) and NOVA
(Netherlands).  SPHERE also received funding from the European
Commission Sixth and Seventh Framework Programmes as part of the
Optical Infrared Coordination Network for Astronomy (OPTICON) under
grant number RII3-Ct-2004-001566 for FP6 (2004–2008), grant number
226604 for FP7 (2009–2012) and grant number 312430 for FP7
(2013–2016). M. B. thank A. Best, K. Allers, G. Mace, E. Artigau,
B. Gauza, R. D. Rosa, M.-E. Naud, F.-R. Lachapelle, J. Patience,
J. Gizis, A. Burgasser, M. Liu, A. Schneider, K. Aller, B. Bowler,
S. Hinkley, and K. Kellogg for providing their spectra of young brown
and companions.  This publication makes use of VOSA, developed under
the Spanish Virtual Observatory project supported from the Spanish
MICINN through grant AyA2011-24052. This research has benefitted from
the SpeX Prism Spectral Libraries, maintained by Adam Burgasser at
http://pono.ucsd.edu/$\sim$adam/browndwarfs/spexprism. We thank 
F. Marocco \& P. Lucas for the generation of the Forsterite reddenning curve used
in Fig. 8. We acknowledge support from the “Progetti Premiali” funding scheme of MIUR. Finally, J.\,O. acknowledges financial support from ICM N\'ucleo Milenio de Formaci\'on Planetaria, NPF. 

\end{acknowledgements}

%\begin{thebibliography}{}

%\bibitem[Alibert et al.(2004)]{ali04} Alibert Y., Mordasini C. \& Benz W. 2004, A\&A, 417, L25

%\end{thebibliography}

\bibliography{References-hd95086}

\Online

\begin{appendix} %First online appendix

\section{Conversion to flux}
\label{app:convflux}

We gather the optical to mid-IR photometry of the star HD\,95086 using
VOSA\footnote{http://svo2.cab.inta-csic.es/theory/vosa/}
\citep{bayo2008}. The measurements were then adjusted by a BT-NEXTGEN synthetic spectrum 
to derive the physical fluxes of HD\,95086 and HD\,95086\,b in each filter and spectral channel used for the characterization of the spectral energy distribution of HD\,95086\,b.

\begin{figure}[h]
\begin{center}
\includegraphics[width=\linewidth]{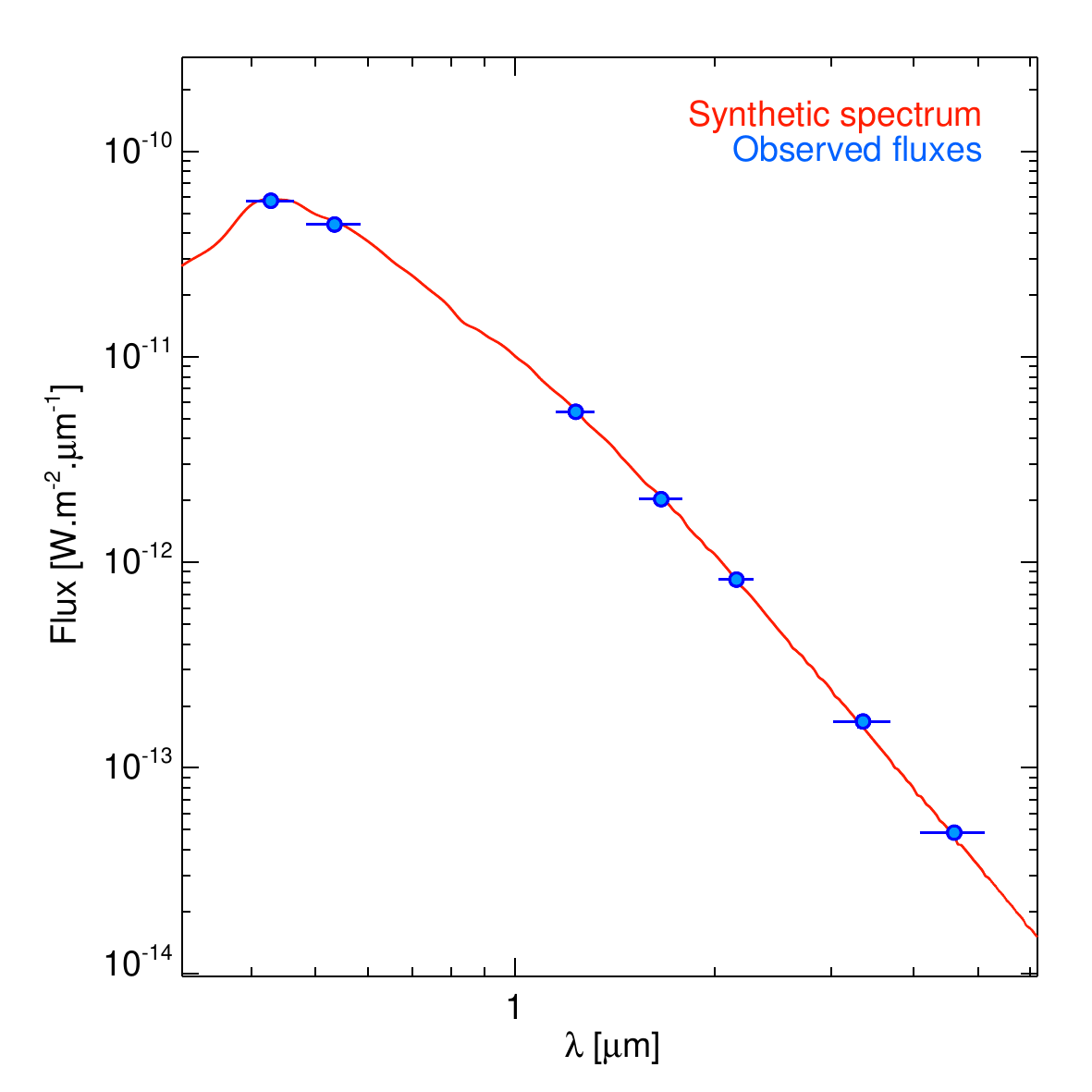}
\caption{BT-NEXTGEN synthetic spectrum (red curve) adjusted onto the star HD\,95086 photometry (blue dots).}
\label{Fig:SEDstar}
\end{center}
\end{figure}

\section{Markov-Chain Monte-Carlo inclination-restricted results}
\label{App:A}

\begin{figure*}[!h]
\hspace{0.2cm}
\includegraphics[width=16cm]{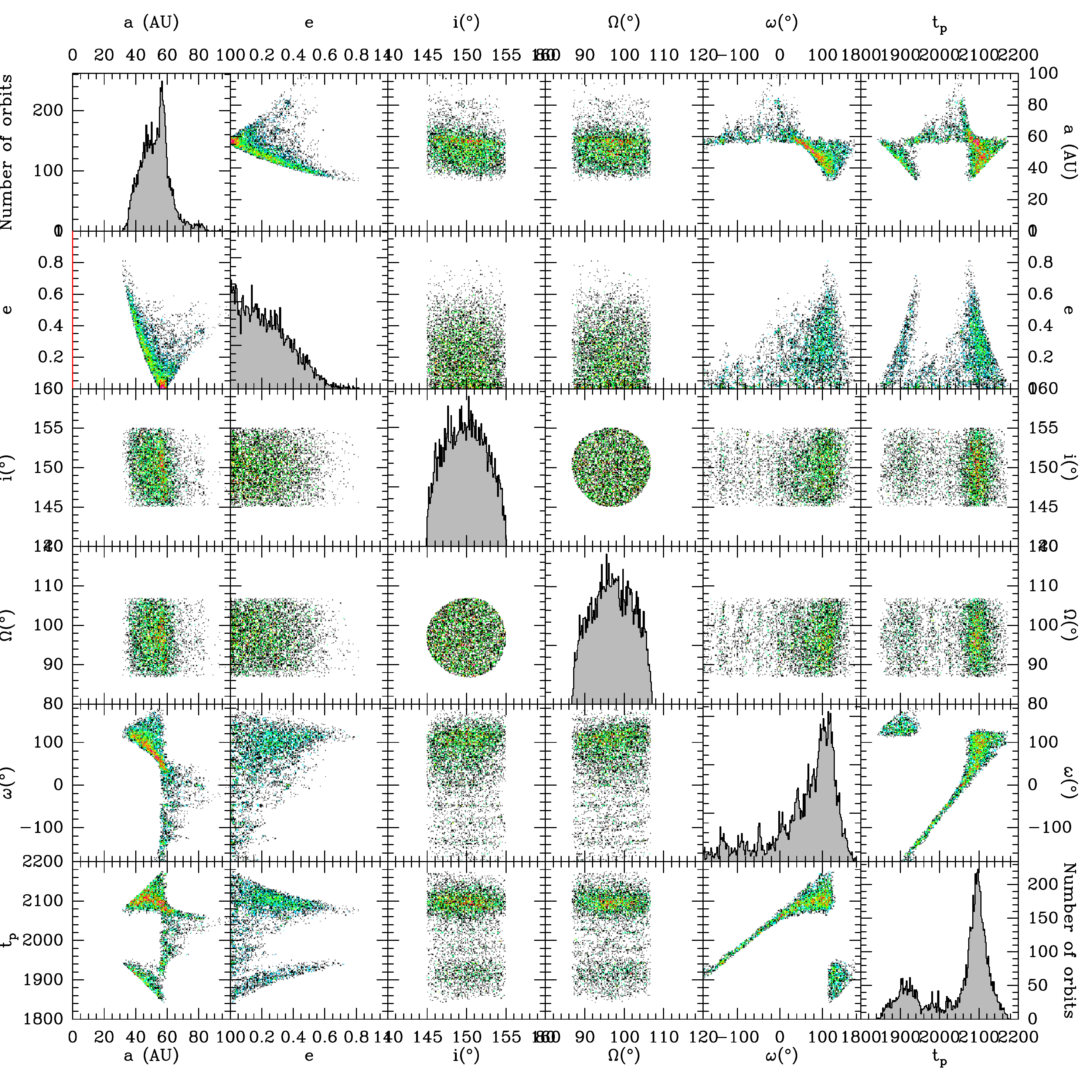} 
\begin{centering}

\caption{Results of the MCMC fit of the NaCo and SPHERE combined astrometric  data of
  HD\,95086\,b reported in terms of statistical distribution matrix of
  the orbital elements $a$, $e$, $i$, $\Omega$, $\omega$ and
  $t_p$. The \textit{red} line indicates the position of the best LSLM
  $\chi_r^2$ model obtained for comparison. The reported solutions
  correspond for the restricted case of orbits with inclination $i$ and longitudes of ascending 
nodes $\Omega$ that ensure a tilt angle with the outer belt resolved with \emph{ALMA} 
($i=150\degr$, $\Omega=97\degr$, \citealt{su2017}) less than $5\degr$.}

\label{fig:astrofull}
\end{centering}
\end{figure*}

\section{Markov-Chain Monte-Carlo results for combined NaCo and GPI measurements}
\label{App:B}

\begin{figure*}[!h]
\hspace{0.2cm}
\includegraphics[width=16cm]{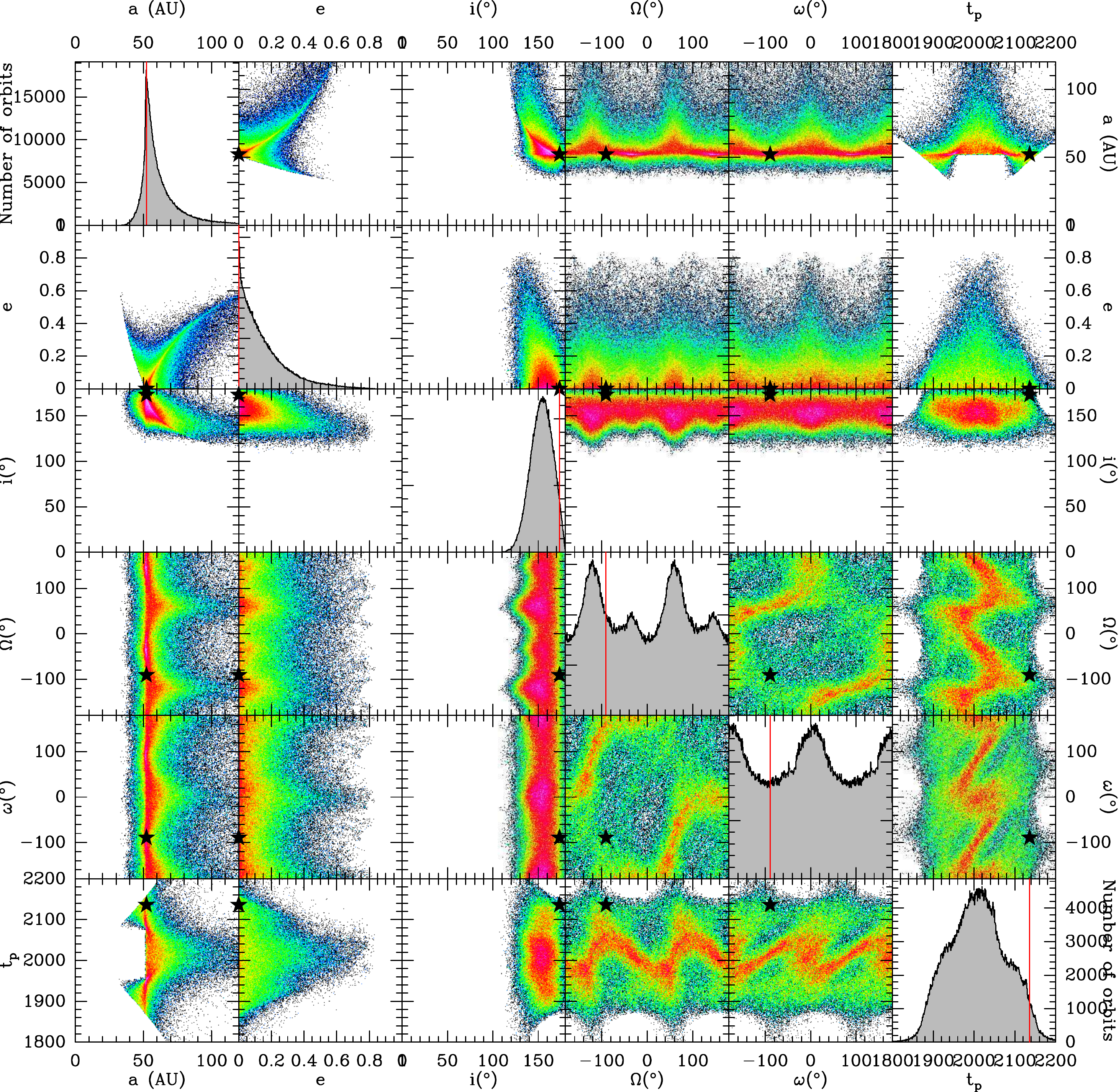} 
\begin{centering}

\caption{Results of the MCMC fit of the NaCo and GPI combined astrometric  data of
  HD\,95086\,b. Measurements are directly taken from Table 1 of \citep{rameau2016}. The results are reported in terms of statistical distribution matrix of
  the orbital elements $a$, $e$, $i$, $\Omega$, $\omega$ and
  $t_p$. The \textit{red} line indicates the position of the best LSLM
  $\chi_r^2$ model obtained for comparison. Considering the new distance estimate of 83.8\,pc, a correction factor of 0.93 in sma and 0.89 in period must be applied for direct comparison with Fig.\,3 of \citep{rameau2016}. The MCMC tool used in this work and the MCMC OFTI tool of \citep{blunt2017} show a relatively good match conforting both MCMC methods.}

\label{fig:astrofull}
\end{centering}
\end{figure*}

%\begin{figure*}[!h]
%\hspace{0.2cm}
%\includegraphics[width=16cm]{Figures/Orbit/MCMC2/mosaique_univ_hd95086_3.pdf}
%\begin{centering}

%\caption{Results of the MCMC fit considering of the combined NaCo and GPI astrometric data of
%  HD\,95086\,b reported in terms of statistical distribution matrix of
%  the orbital elements $q$, $e$, $i$, $\Omega$, $\omega$ and
%  $t_p$. The \textit{red} line indicates the position of the best LSLM
%  $\chi_r^2$ model obtained for comparison. }

%\label{fig:astrofull}
%\end{centering}
%\end{figure*}

%\section{Markov-Chain Monte-Carlo results considering the NaCo and GPI astrometry}

%% \begin{figure*}[pth]
%% \hspace{0.2cm}
%% \includegraphics[width=18cm]{Markov-Chain Monte-Carlo orbital fitting}
%% \begin{centering}
%% \caption{LSMC}
%% \label{fig:astrofull}
%% \end{centering}
%% \end{figure*}
\section{Benchmark dusty L dwarfs}
\label{App:C}

\begin{table*}
\caption{Properties of  dusty, peculiar, and possibly young, dwarfs later than L4 considered in our study}
\label{tab:AppA}
\begin{center}
\begin{tabular}{lllll}
\hline\hline
Source name								&		Spectral type		&				Membership				&	Mass   &	Reference \\
												&								&										&	($\mathrm{M_{Jup}}$)		& \\
\hline
2MASS J11193254-1137466		&		L7			&		  TWA				 &	4.3-7.6		&	1, 2	\\
2MASS J17081563+2557474	&	L5		&	\dots				&		\dots		&	1 \\
WISEP J004701.06+680352.1 & L7	&	AB Dor 	 &	$\sim$18  &	3, 4 \\
PSO J318.5338-22.8603	&	L7	&	$\beta$ Pic		&	$8.3\pm0.5$	&	5, 6 \\
ULAS J222711-004547 & L7	&	\dots & \dots & 7 \\
WISE J174102.78-464225.5 & L7 & $\beta$ Pic or AB Dor & 4-21 & 8 \\
WISE J020625.27+264023.6 & L8 & \dots & \dots & 9, 22 \\
WISE J164715.57+563208.3 & L9 & Argus  & 4-5 & 9, 10 \\
2MASS J00011217+1535355	&	L4	& AB Dor	&	$25.3 \pm 1.0$	&	 11, 12, 13 \\
2MASS J21543454-1055308 & L4	& Argus?	&	\dots & 14, 22 \\
2MASS J22064498-4217208	&	L4	&	AB Dor	&	$23.1\pm6.4$ & 15 \\ 
2MASS J23433470-3646021 & L3-L6	& AB Dor	&	\dots & 12 \\
2MASS J03552337+113343 & L5	&	AB Dor & 13-30 & 16, 17, 18 \\
2MASS J21481628+4003593  & L6	&	Argus?	& \dots	&		10, 15, 19, 22\\
2MASSW J2244316+204343 & L6	&	AB Dor	&	11-12	&	15, 20 \\
2MASSJ05012406–0010452 & L4	& Columba	& $10.2^{+0.8}_{-1.0}$ & 12, 21 \\
\hline
\end{tabular}
\end{center}
\tablefoot{References: 1-\cite{2015AJ....150..182K}, 2-\cite{2016ApJ...821L..15K}, 3-\cite{2012AJ....144...94G}, 4-\cite{2015ApJ...799..203G}, 5-\cite{2013ApJ...777L..20L}, 6-\cite{2016ApJ...819..133A}, 7-\cite{2014MNRAS.439..372M}, 8-\cite{2014AJ....147...34S}, 9-\cite{2011ApJS..197...19K}, 10-\cite{2014ApJ...783..121G}, 11-\cite{2004AJ....127.3553K}, 12-\cite{2015ApJS..219...33G}, 13-\cite{2015ApJ...798...73G}, 14-\cite{2014ApJ...792L..17G}, 15-\cite{2016ApJS..225...10F}, 16-\cite{2006AJ....132..891R}, 17-\cite{2009AJ....137.3345C}, 18-\cite{2013AJ....145....2F}, 19-\cite{2008ApJ...686..528L}, 20-\cite{2003ApJ...596..561M}, 21-\cite{2008MNRAS.390.1517C}, 22-\cite{2016ApJ...833...96L}}\\
%\tablefoot{Distances: $^{k}=$kinematic distance,  $^{\pi}$=parallax,  $^{p}$=photometric.}\\
\end{table*}

\end{appendix}

\end{document}